\documentclass[16pt]{article}
\usepackage{amscd}
\usepackage{bbm}
\usepackage{mathrsfs}
\usepackage{amssymb}
\usepackage{graphics}
\usepackage{graphicx}
\usepackage{amsfonts}
\usepackage{amsmath}
\usepackage{array}
\usepackage{booktabs}
\usepackage{cite}

\usepackage[numbers,sort&compress]{natbib}
\begin{document}
\date{}
\title{The co-directional overtaking collision problem between a dispersive shock wave and a rarefaction wave for the Hirota equation}
\author{Yuan Xiang, Rui Guo$\thanks{Corresponding author:
guorui@tyut.edu.cn}$ and Hua-Ying Ren\
\\
\\{\em
School of Mathematics, Taiyuan University  of} \\
{\em Technology, Taiyuan 030024, China}
} \maketitle

\begin{abstract}
In this paper, we mainly investigate the overtaking collision problem between a dispersive shock wave (DSW) and a rarefaction wave (RW) propagating in the same direction in the defocusing Hirota equation framework. Based on the Whitham theory, the zero-phase and single-phase modulation systems corresponding to the defocusing Hirota equation are given, along with explicit expressions of Riemann invariants. For step-type initial conditions with left-side DSW and right-side RW propagating identically, the global initial configuration is constructed.~The modulation solution within the nonlinear collision domain is obtained via the generalized hodograph transformation and solutions of the Euler-Poisson-Darboux equation.~Thus, by matching the Riemann invariants at the boundaries of DSW and RW, we can provide a detailed analytical description for the dynamical behaviors of the collision across all evolutionary stages.~The accuracy of theoretical predictions is verified by comparison with direct numerical simulations.

\vspace{5mm}\noindent\emph{Keywords}: Whitham theory; Dispersive shock wave; Rarefaction wave; Co-directional overtaking collision; Hirota equation
\end{abstract}
\vspace{7mm}\noindent\textbf{1  Introduction}
\hspace*{\parindent}
\renewcommand{\theequation}{\arabic{equation}}\\

Nonlinear wave phenomena are ubiquitous in nature and engineering applications, ranging from surface fluctuations in fluids~\cite{ck1,ck2,ck3} and ion acoustic waves in plasmas~\cite{ck4,ck5}, to optical pulse propagation in nonlinear optics~\cite{ck6,ck7} and hydrodynamic wave behaviors in Bose-Einstein condensates (BECs) of ultracold atomic systems~\cite{ck8,ck9,ck10,ck11}. Two typical types of nonlinear waves exist in nonlinear systems: rarefaction waves (RWs) and dispersive shock waves (DSWs). A RW forms due to nonlinear expansion, featuring a monotonic profile without oscillations~\cite{ck1,ck12}. A DSW arises from the interplay between nonlinear steepening and dispersion, whereby dispersion prevents wave breaking and eventually shapes a wave train with damped oscillations~\cite{ck1,ck13,ck14}. The generation mechanisms, propagation properties and interaction rules of these waves stand as core research topics in nonlinear science and dispersive fluid mechanics.

DSWs achieve smooth transitions between different uniform states through periodic oscillatory wave trains and function as dispersion-regularized structures that replace classical dissipation-regularized viscous shock waves (VSWs)~\cite{ck14,ck15}. Early studies on shock waves mainly focused on viscous dissipation. Since Gurevich and Pitaevskii pioneered the application of the Whitham modulation theory to describe DSWs in the Korteweg-de Vries (KdV) equation back in the 1960s~\cite{ck16}, researchers have been able to systematically analyze nonlinear wave dynamics dominated by dispersion. Within the framework of Gurevich-Pitaevskii theory, Whitham modulation theory incorporates the amplitude, frequency and wavenumber of wave trains into a set of unified evolution equations, and derives modulation equations that govern the formation and evolution of DSWs. For integrable systems, these modulation equations can be rewritten in the diagonal form of Riemann invariants, which greatly simplifies theoretical analysis.
In quantum fluid systems such as BECs, the Gross-Pitaevskii (GP) equation~\cite{ck17,ck18} accurately describes the macroscopic wave dynamics of condensates~\cite{ck9,ck10}. Under the fluid approximation, it reduces to a shallow-water-like equation, which provides a fundamental model for investigating the generation, propagation and interaction of DSWs and RWs. So far, the evolutionary laws of isolated DSWs and RWs have been well explored. Using the KdV~\cite{ck19,ck20}, nonlinear Schrödinger (NLS)~\cite{ck21,ck22} and modified KdV (mKdV) equations~\cite{ck23,ck24,ck25}, scholars have revealed key properties including the self-similar evolution of single-phase DSWs, self-similar expansion of RWs and decomposition of Riemann invariants. These findings build a consistent understanding connecting integrable systems to experimental platforms.

On this basis, interactions among multiple nonlinear waves are far more complicated and carry richer physical implications, gradually becoming a major research hotspot. Among all interaction scenarios, the interaction between DSWs and RWs is of particular importance. Existing theoretical and numerical results indicate that dispersive systems have no entropy generation, so DSW-RW interactions exhibit distinctive behaviors. For instance, in the KdV equation, when a DSW overtakes a RW propagating in the same direction, solitons may escape from the interaction region, or a low-amplitude wave train decaying over time may form in the wave tail~\cite{ck26,ck27}. In the defocusing NLS equation, head-on collisions between DSWs and RWs lead to waveform refraction~\cite{ck28,ck29,ck30}. Unlike classical VSWs, dispersion-dominated wave interactions involve negligible energy dissipation and no waveform annihilation~\cite{ck31,ck32,ck33}. Both waves retain their core structures after collision, demonstrating the essential features of dispersive conservation and elastic interaction. In BEC experiments, researchers have observed head-on collisions, boundary refraction of DSWs and RWs, which offer direct experimental evidence for relevant theoretical models~\cite{ck28}.

Although DSW-RW interactions in the NLS and KdV equations have been systematically studied, higher-order dispersive corrections are often required to precisely characterize wave propagation in realistic physical systems. Third-order dispersion cannot be neglected in scenarios such as deep-water wave dynamics and ultrashort optical pulse transmission. Under such circumstances, the Hirota equation~\cite{ck34} becomes a representative model:
\begin{equation}\label{1}
i \varepsilon \psi_{\mathrm{\it t}}+\alpha \left(\varepsilon^2 \psi_{\mathrm{\it xx}}-2\left| \psi \right|^2\psi \right) +i\beta \left(\varepsilon^3 \psi_{\mathrm{\it xxx}}-6 \varepsilon \left| \psi \right|^2\psi_{\mathrm{\it x}} \right) =0,
\end{equation}
where $\beta$ denotes the third-order dispersion coefficient. As a member of the integrable NLS hierarchy, the Hirota equation possesses a Lax pair structure, making it applicable for analysis via the finite-gap integration theory and Whitham modulation method. Compared with the standard NLS equation, the higher-order dispersive term in the Hirota equation significantly modifies the dispersion relation of DSWs, the explicit form of Riemann invariants, and the interaction rules between different waves. The Riemann problem and general cubic wave breaking problem for the Hirota equation have been thoroughly investigated~\cite{ck35,ck36}. Nevertheless, critical issues concerning overtaking collisions, including waveform evolution, modulation intensity variation and self-similar asymptotic solutions, remain unresolved.

Building on the established theories of zero-phase and single-phase Whitham modulation, this paper systematically investigates the full dynamical behaviors of co-propagating overtaking collisions between DSWs and RWs in Eq.~(\ref{1}). We first construct global initial profiles for the step-type initial condition, where a DSW is placed on the left and a RW on the right, with both waves traveling in the same direction. Furthermore, we adopt the generalized hodograph transformation~\cite{ck37} to analyze waveform evolution, velocity variation and boundary motion at different stages of the interaction region. We also compare our theoretical predictions with direct numerical simulations implemented by the pseudo-spectral method~\cite{ck38} to verify accuracy. This work improves the general theory of nonlinear wave interactions, and provides theoretical references for experiments on superfluids, nonlinear optics, water waves and related fields.

This paper is structured as follows. In Section~$2$, we review the modulation theory for the defocusing Hirota equation, covering zero-phase Whitham equations, single-phase Whitham equations and periodic solutions. In Section~$3$, we construct global initial data and derive the Euler-Poisson-Darboux equation using the generalized hodograph transformation. In Section~$4$, we analyze all stages of the overtaking collision, with detailed discussions on subregions of the DSW-RW interaction zone. In Section~$5$, we summarize the main conclusions and present prospects for future research.

\vspace{5mm} \noindent\textbf{2  Fundamentals of modulation theory in the defocusing Hirota equation }
\hspace*{\parindent}
\renewcommand{\theequation}{2.\arabic{equation}}\setcounter{equation}{0}\\

In this section,  we mainly present relevant results of the Whitham slow‑varying modulation theory for nonlinear waves within the framework of Eq.~(\ref{1}), which provide an essential foundation for analyzing the interaction between a DSW and a RW in subsequent sections. For the research presented herein, the Whitham slow-varying modulation theory, categorized into two fundamental types (zero-phase modulation and single-phase modulation), fully describes the spatiotemporal evolution of co-direction overtaking collisions. The first derivation of the zero-phase and single-phase modulation equations for Eq.~(\ref{1}) was reported in the Ref.~\cite{ck35}, achieved by means of the finite-gap integration method and averaged conservation laws~\cite{ck35,ck36,ck39}.
\vspace{5mm}\\
\textbf{2.1 The zero-phase modulation equations}
\\

The Eq.~(\ref{1}) can be converted to the form of dispersive hydrodynamics by introducing the Madelung transformation
\begin{equation}\label{2.1}
\psi\left( x,t \right) =\sqrt{\rho \left( x,t \right)}e^{\frac{i}{\varepsilon}\phi},\;\;\;\;\phi _{\mathrm{\it x}}=u\left( x,t \right)
\end{equation}
and it becomes
\begin{equation}\label{2.2}
\begin{aligned}
\begin{split}
&\rho _{\mathrm{\it t}}+\left( 2\alpha \left( \rho u \right) -3\beta \left( \rho ^2+u^2\rho \right) \right) _{\mathrm{\it x}}=-\varepsilon^2 \beta \left( 4\rho ^{3/4}\left( \rho ^{1/4} \right) _{\mathrm{\it xx}} \right) _{\mathrm{\it x}},
\\
&u_{\mathrm{\it t}}+\left( 2\alpha \left( \rho +\frac{1}{2}u^2 \right) -\beta \left( 6u\rho +u^3 \right) \right) _{\mathrm{\it x}}=\varepsilon^2 \left( \frac{1}{2}\alpha \left( \frac{\rho _{\mathrm{\it xx}}}{\rho}-\frac{\rho _{\mathrm{\it x}}^{2}}{2\rho ^2} \right) +\beta \left( \frac{3u\rho _{\mathrm{\it x}}^{2}}{4\rho ^2}-\frac{3u\rho _{\mathrm{\it x}}}{2\rho}-\frac{3u\rho _{\mathrm{\it xx}}}{2\rho}-u_{\mathrm{\it xx}} \right) \right) _{\mathrm{\it x}},
\end{split}
\end{aligned}
\end{equation}
where $\rho \left( x,t \right) > 0$ and $u \left( x,t \right)$ are real-valued functions, which represent density and flow velocity of the hydrodynamics, respectively.

By setting $\varepsilon = 0$, the system~(\ref{2.2}) can become the following dispersionless form
\begin{equation}\label{2.3}
\begin{aligned}
\begin{split}
&\rho _{\mathrm{\it t}}+\left( 2\alpha \rho u  -3\beta \left( \rho ^2+u^2\rho \right) \right) _{\mathrm{\it x}}=0,
\\
&u_{\mathrm{\it t}}+\left( 2\alpha \left( \rho +\frac{1}{2}u^2 \right) -\beta \left( 6u\rho +u^3 \right) \right) _{\mathrm{\it x}}=0,
\end{split}
\end{aligned}
\end{equation}
which can be transformed to a diagonal form, that is to say, the zero-phase Whitham equations 
\begin{equation}\label{2.4}
\frac{\partial \lambda _+}{\partial t}+v_+\frac{\partial \lambda _+}{\partial x}=0,
\;\;\;\;
\frac{\partial \lambda _-}{\partial t}+v_-\frac{\partial \lambda _-}{\partial x}=0,
\end{equation}
by introducing the Riemann invariants
\begin{equation}\label{2.5}
\lambda _{\pm}=-u/2\pm \sqrt{\rho},
\end{equation}
where the characteristic velocities are
\begin{equation}\label{2.6}
\begin{aligned}
\begin{split}
&v_+=-\alpha \left( 3\lambda _++\lambda _- \right) -\frac{3}{2}\beta \left( 5\lambda _{+}^{2}+2\lambda _+\lambda _-+\lambda _{-}^{2} \right) ,
\\
&v_-=-\alpha \left( 3\lambda _-+\lambda _+ \right) -\frac{3}{2}\beta \left( 5\lambda _{-}^{2}+2\lambda _+\lambda _-+\lambda _{+}^{2} \right) .
\end{split}
\end{aligned}
\end{equation}
Moreover, $\rho$ and $u$ can be expressed in terms of $\lambda_{\pm}$ by the formulas
\begin{equation}\label{2.7}
\rho =\frac{1}{4}\left( \lambda _+-\lambda _- \right) ^2,\;\;\;\;u=-\lambda _+-\lambda _-.
\end{equation}

The Eqs.~(\ref{2.4}) correspond to a uniform steady wave field free of periodic oscillations, which can be used to characterize the background flow field far from the wave interaction region. In this state, physical parameters such as wave density and flow velocity remain constant throughout the domain, with no spatiotemporal gradients or deformational evolution, representing the fundamental equilibrium state of the wave system. Furthermore, the Eqs.~(\ref{2.4}) are particularly suited for characterizing RWs in defocusing nonlinear waves, and act as the fundamental governing equations for nonlinear waves free of shocks and with self-similar expansion.
\vspace{5mm}\\
\textbf{2.2 The single-phase modulation equations}
\\

We review briefly the periodic solution of Eq.~(\ref{1}) in the framework of the AKNS system, and calculate the two special limits regarding the solution (more details can be found in Refs.~\cite{ck36,ck37}). It can be expressed in terms of the Jacobi elliptic sine function sn:
\begin{equation}\label{2.8}
\rho \left( x,t \right) =\frac{1}{4}\left( \lambda _1-\lambda _2-\lambda _3+\lambda _4 \right) ^2+\left( \lambda _1-\lambda _2 \right) \left( \lambda _3-\lambda _4 \right) \mathrm{sn}^2\left( \sqrt{\left( \lambda _1-\lambda _3 \right) \left( \lambda _2-\lambda _4 \right)}\left(\frac{x-Vt-Q}{\varepsilon} \right) ,m \right) ,
\end{equation}
where $\lambda _1\leqslant \lambda _2\leqslant \lambda _3\leqslant \lambda _4$ are four free parameters, the phase velocity $V$ and  modulus $0\leqslant m\leqslant 1$ of the  elliptic function equal to
\begin{equation}\label{2.9}
V=-\alpha \sum_{i=1}^4{\lambda _{\mathrm{\it i}}}-\beta \left( \frac{3}{2}\left( \sum_{i=1}^4{\lambda _{\mathrm{\it i}}} \right) ^2-2\sum_{1\leqslant i\leqslant j\leqslant 4}{\lambda _{\mathrm{\it i}}\lambda _{\mathrm{\it j}}} \right) ,\;\;\;\;m=\frac{\left( \lambda _1-\lambda _2 \right) \left( \lambda _3-\lambda _4 \right)}{\left( \lambda _1-\lambda _3 \right) \left( \lambda _2-\lambda _4 \right)}.
\end{equation}

The wavelength $L$ can be expressed through the complete elliptic integral of the first kind $K\left( m \right) $ as
\begin{equation}\label{2.10}
L=\frac{2K\left( m \right)}{\sqrt{\left( \lambda _1-\lambda _3 \right) \left( \lambda _4-\lambda _2 \right)}}.
\end{equation}
In the limit $\lambda _3\rightarrow \lambda _2$ and $m\rightarrow 1,$ the wavelength tends to infinity and the periodic solution~(\ref{2.8}) becomes a dark soliton solution
\begin{equation}\label{2.11}
\rho \left( x,t \right) =\rho_{\mathrm{\it s}}-\left( \lambda _1-\lambda _2 \right) \left( \lambda _2-\lambda _4 \right) \mathrm{sech} ^2\left( \sqrt{\left( \lambda _1-\lambda _2 \right) \left( \lambda _2-\lambda _4 \right)}\left( x-V_{\mathrm{\it s}}t-Q \right) \right) ,
\end{equation}
where the background density $\rho_{\mathrm{\it s}}$ and velocity $V_{\mathrm{\it s}}$ are expressed with $\lambda_1, \lambda_2,\lambda_4$ as
\begin{equation}\label{2.12}
\rho_{\mathrm{\it s}} = \frac{1}{4}\left( \lambda _1-\lambda _4 \right) ^2, V_{\mathrm{\it s}}=-\alpha \left( \lambda _1+2\lambda _2+\lambda _4 \right) -\beta \left( \frac{3}{2}\left( \lambda _1+2\lambda _2+\lambda _4 \right) ^2-2\left( \lambda _{2}^{2}+\lambda _1\lambda _4+2\lambda _1\lambda _2+2\lambda _2\lambda _4 \right) \right) .
\end{equation}

In the opposite limit $m\rightarrow 0\left( \lambda _3\rightarrow \lambda _4\,\,or\,\,\lambda _2\rightarrow \lambda _1 \right) ,$ and the periodic solution~(\ref{2.8}) becomes a small-amplitude linear harmonic wave
\begin{equation}\label{2.13}
\rho \left( x,t \right) =\frac{1}{4}\left( \lambda _1-\lambda _2 \right) ^2,\; \text{or}\;\;\rho \left( x,t \right) =\frac{1}{4}\left( \lambda _3-\lambda _4 \right) ^2.
\end{equation}

On this basis, we know that within the framework of Gurevich-Pitaevskii theory, this periodic solution~(\ref{2.8}) needs to be modulated by the Whitham equations to serve as an approximate solution for the DSW, i.e., $\lambda_{\mathrm{\it i}}$ are slowly-varying functions of $x$ and $t$~\cite{ck16}. Accordingly, the Whitham equations corresponding to the Riemann invariants $\lambda_{\mathrm{\it i}}$ are given as follows:
\begin{equation}\label{2.14}
\frac{\partial \lambda _{\mathrm{\it i}}}{\partial t}+v_{\mathrm{\it i}}\left( \lambda _1,\lambda _2,\lambda _3,\lambda _4 \right) \frac{\partial \lambda _{\mathrm{\it i}}}{\partial x}=0,\;\;\;\;i=1,2,3,4,
\end{equation}
where the Whitham velocities are 
\begin{equation}\label{2.15}
v_{\mathrm{\it i}}\left( \lambda _1,\lambda _2,\lambda _3,\lambda _4 \right) =\left( 1-\frac{L}{\partial _{\lambda _{\mathrm{\it i}}}L}\partial _{\lambda _{\mathrm{\it i}}} \right) V.
\end{equation}
Substituting Eqs.~(\ref{2.15}) into~(\ref{2.14}) yields the explicit expressions for the characteristic velocities
\begin{equation}\label{2.16}
\begin{aligned}
\begin{split}
&v_1=V-2\left[ \alpha +\beta \left( 2\lambda _1+\sum_{i=1}^4{\lambda _{\mathrm{\it i}}} \right) \right] \frac{\left( \lambda _1-\lambda _2 \right) \left( \lambda _1-\lambda _4 \right)}{\left( \lambda _1-\lambda _4 \right) +\left( \lambda _4-\lambda _2 \right) \kappa(m)},
\\
&v_2=V+2\left[ \alpha +\beta \left( 2\lambda _2+\sum_{i=1}^4{\lambda _{\mathrm{\it i}}} \right) \right] \frac{\left( \lambda _1-\lambda _2 \right) \left( \lambda _2-\lambda _3 \right)}{\left( \lambda _2-\lambda _3 \right) +\left( \lambda _3-\lambda _1 \right) \kappa(m)},
\\
&v_3=V-2\left[ \alpha +\beta \left( 2\lambda _3+\sum_{i=1}^4{\lambda _{\mathrm{\it i}}} \right) \right] \frac{\left( \lambda _2-\lambda _3 \right) \left( \lambda _3-\lambda _4 \right)}{\left( \lambda _2-\lambda _3 \right) -\left( \lambda _2-\lambda _4 \right) \kappa(m)},
\\
&v_4=V+2\left[ \alpha +\beta \left( 2\lambda _4+\sum_{i=1}^4{\lambda _{\mathrm{\it i}}} \right) \right] \frac{\left( \lambda _1-\lambda _4 \right) \left( \lambda _3-\lambda _4 \right)}{\left( \lambda _1-\lambda _4 \right) -\left( \lambda _1-\lambda _3 \right) \kappa(m)},
\end{split}
\end{aligned}
\end{equation}
where $\kappa(m)=\frac{E(m)}{K(m)}$.

The formulations for the velocities $v_i$ in different limit are presented as follows. In the soliton limit $m\rightarrow 1\left( \lambda _3\rightarrow \lambda _2 \right) $, the Whitham velocities reduce to
\begin{equation}\label{2.17}
\begin{aligned}
\begin{split}
&v_1=-\alpha \left( 3\lambda _1+\lambda _4 \right) -\frac{3}{2}\beta \left( 5\lambda _{1}^{2}+2\lambda _1\lambda _4+\lambda _{4}^{2} \right) ,
\\
&v_2=v_3=-\alpha \left( \lambda _1+2\lambda _2+\lambda _4 \right) -\beta \left( \frac{3}{2}\left( \lambda _{1}^{2}+\lambda _{4}^{2} \right) +\left( \lambda _1+2\lambda _2 \right) \left( 2\lambda _2+\lambda _4 \right) \right) ,
\\
&v_4=-\alpha \left( 3\lambda _4+\lambda _1 \right) -\frac{3}{2}\beta \left( 5\lambda _{4}^{2}+2\lambda _1\lambda _4+\lambda _{1}^{2} \right) .
\end{split}
\end{aligned}
\end{equation}
Analogously, in the harmonic limit $m\rightarrow 0\left( \lambda _2\rightarrow \lambda _1 \right) $, the Whitham velocities reduce to
\begin{equation}\label{2.18}
\begin{aligned}
\begin{split}
&v_1=v_2=-\alpha \left( 4\lambda _2-\frac{\left( \lambda _3-\lambda _4 \right) ^2}{2\lambda _2-\lambda _3-\lambda _4} \right) -\beta \left( -12\lambda _{1}^{2}-\frac{3}{2}\left( \lambda _3-\lambda _4 \right) ^2\frac{\lambda _3+\lambda _4+2\lambda _1}{\lambda _3+\lambda _4-2\lambda _1} \right) ,
\\
&v_3=-\alpha \left( 3\lambda _3+\lambda _4 \right) -\frac{3}{2}\beta \left( 5\lambda _{3}^{2}+2\lambda _3\lambda _4+\lambda _{4}^{2} \right) ,
\\
&v_4=-\alpha \left( 3\lambda _4+\lambda _3 \right) -\frac{3}{2}\beta \left( 5\lambda _{4}^{2}+2\lambda _3\lambda _4+\lambda _{3}^{2} \right) .
\end{split}
\end{aligned}
\end{equation}
In another harmonic limit $m\rightarrow 0\left( \lambda _3\rightarrow \lambda _4 \right) $, the Whitham velocities become
\begin{equation}\label{2.19}
\begin{aligned}
\begin{split}
&v_1=-\alpha \left( 3\lambda _1+\lambda _2 \right) -\frac{3}{2}\beta \left( 5\lambda _{1}^{2}+2\lambda _1\lambda _2+\lambda _{2}^{2} \right) ,
\\
&v_2=-\alpha \left( 3\lambda _2+\lambda _1 \right) -\frac{3}{2}\beta \left( 5\lambda _{2}^{2}+2\lambda _1\lambda _2+\lambda _{1}^{2} \right) ,
\\
&v_3=v_4=-\alpha \left( 4\lambda _2-\frac{\left( \lambda _1-\lambda _2 \right) ^2}{2\lambda _4-\lambda _1-\lambda _2} \right) -\beta \left( -12\lambda _{4}^{2}-\frac{3}{2}\left( \lambda _1-\lambda _2 \right) ^2\frac{\lambda _1+\lambda _2+2\lambda _4}{\lambda _1+\lambda _2-2\lambda _4} \right) .
\end{split}
\end{aligned}
\end{equation}

To characterize DSWs, we adopt the Gurevich-Pitaevskii boundary matching scheme and impose appropriate initial and boundary conditions on the Riemann invariants $\lambda_i$ to constrain the Whitham system, thereby establishing the corresponding definite conditions for initially right-traveling DSWs. Specifically, within the framework of Gurevich-Pitaevskii theory, the upper half-plane is divided into three regions by the leading edge and trailing edge of the DSW: the left side of the leading edge, the right side of the trailing edge, and the DSW region. In the outer regions of the leading and trailing edges, the flow obeys the dispersionless limit equation corresponding to the shallow water-like system~(\ref{2.3}) with Riemann invariants $\lambda_\pm$. In the interior region of the DSW, the averaged oscillatory flow is governed by the four Whitham system~(\ref{2.14}) with $\lambda_i$ as the unknown variables. As mentioned above, It can be seen that  $v_{\pm}$ from Eqs.~(\ref{2.4}) coincide with the Whitham velocities corresponding to the two Riemann invariants in the degenerate case of Eqs.~(\ref{2.16}), which facilitates free boundary matching for the global formulation of initial configurations.

\vspace{5mm} \noindent\textbf {\textbf{3  The co-directional overtaking collision problem}}
\hspace*{\parindent}
\renewcommand{\theequation}{3.\arabic{equation}}\setcounter{equation}{0}\\

A DSW and a RW propagate unidirectionally toward the right. Driven by the discrepancy in their respective propagation velocities, the DSW continuously approaches the RW and achieves complete separation subsequent to the collision event. The entire evolutionary process of the collision strictly adheres to the comprehensive dynamical behaviors, which include the global stepped initial configuration, boundary matching of Riemann invariants, and staged spatiotemporal evolution. Consequently, the present section focuses on elaborating the initial configuration, core solution methodologies and the modulation phase shift for the problem of co-direction overtaking collision between a DSW and a RW.
\vspace{5mm}\\
\textbf{3.1 Global stepped zero-phase initial configuration}
\\

To study the overtaking collision between a DSW and a RW propagating in the same direction, we consider the following configuration. At time $t = t_c$, a simple right‑propagating DSW centered at $(0,0)$ expands over the domain $x_1^l(t) < x < x_1^r(t)$, subject to the following transition conditions:
\begin{equation}\label{3.1}
\lambda_{-}\big(x_1^l,t_c\big)=\lambda_{-}\big(x_1^r,t_c\big)=0.
\end{equation}
Meanwhile, a right‑propagating RW centered at $(l,0)$ is located near $x=l$ within the region $x_2^l(t) < x < x_2^r(t)$. In this region, the Riemann invariant satisfies
\begin{equation}\label{3.2}
\lambda_{+}\big(x_2^l,t_c\big)=\lambda_{+}\big(x_2^r,t_c\big)=1.
\end{equation}
These two nonlinear waves are separated by a flat plateau $x_1^r(t) < x < x_2^l(t)$, where the background state is given by $\rho=\dfrac{1}{4}$ and $v=-1$. Consequently, we define $x_1^{l,r}=s_1^{l,r}t$ and $x_2^{l,r}=s_2^{l,r}t+l$, with $s_1^{l,r},s_2^{l,r}>0$ denoting the rightward‑propagating edge velocities of the DSW and the RW, respectively.

Combined with the transition conditions~(\ref{3.1}) and~(\ref{3.2}) for two independent simple waves, the initial conditions of global Riemann invariants are constructed based on distinct polarity jumps of shallow-water Riemann invariants $\lambda_{\pm}$ varying with staggered position $l$ (see Fig.~$1$(a)):
\begin{equation}\label{3.3}
\lambda_{+}(x,0)=
\begin{cases}
h_1, & x<0,\\
1, & x>0;
\end{cases}
\quad \text{and} \quad
\lambda_{-}(x,0)=
\begin{cases}
0, & x<l,\\
h_2, & x>l,
\end{cases}
\end{equation}
where $h_1>1$ and $0<h_2<1$.

The initial density and velocity profiles $\rho(x,0)$ and $v(x,0)$ corresponding to the initial Riemann invariants~(\ref{3.3}) can be acquired using expressions~(\ref{2.5}) in the form of piecewise constant distributions(see Fig.~$1$(b) and (c))
\begin{equation}\label{3.4}
\rho(x,0)=
\begin{cases}
\dfrac{1}{4}h_1^2, & x<0,\\[4pt]
\dfrac{1}{4}, & 0\le x\le l,\\[4pt]
\dfrac{1}{4}(1-h_2)^2, & x>l;
\end{cases}
\quad \text{and} \quad
v(x,0)=
\begin{cases}
-h_1, & x<0,\\
-1, & 0\le x\le l,\\
-(1+h_2), & x>l.
\end{cases}
\end{equation}

\begin{center}
\includegraphics[scale=0.22]{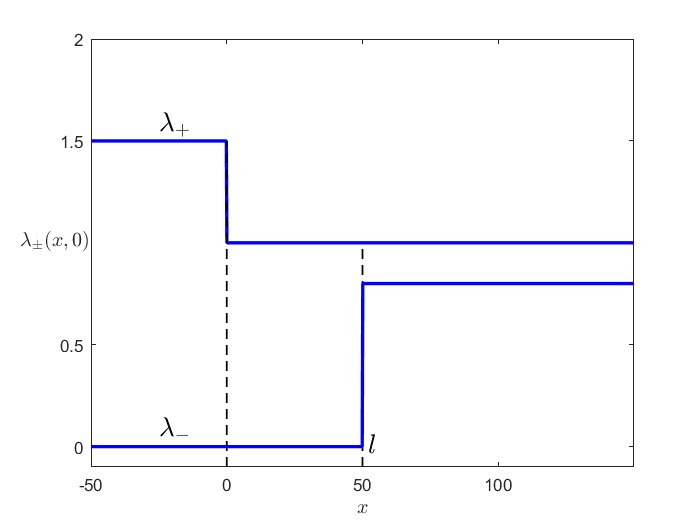}
\includegraphics[scale=0.22]{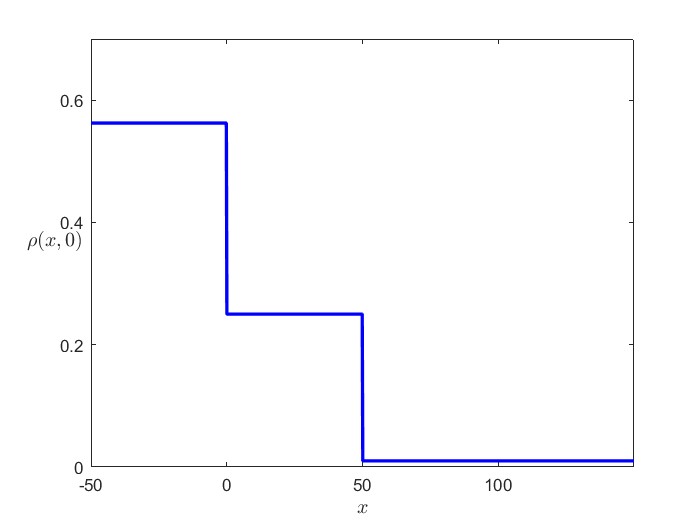}
\includegraphics[scale=0.22]{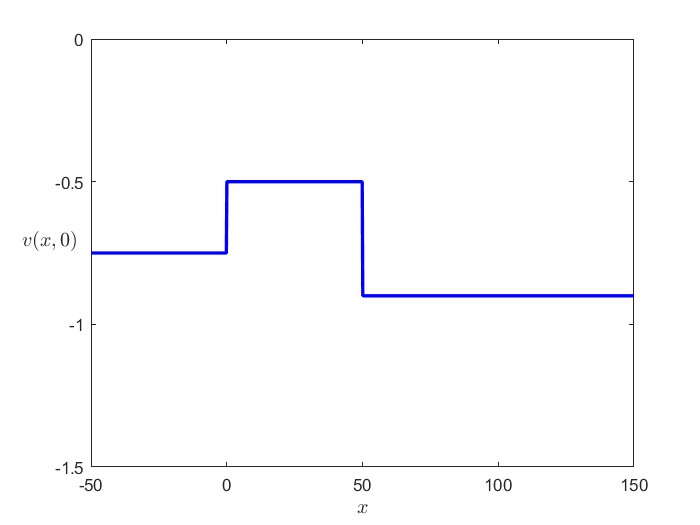}

\vspace{-0.3cm}
{\footnotesize\hspace{0.2cm}(a)\hspace{5.2cm}(b)\hspace{5.2cm}(c)}

\vspace{-0.2cm}
\flushleft{\footnotesize
\textbf{Fig.~$1$.} Global initial configuration of Eqs.~(\ref{2.3}) associated with the co-directional overtaking collision between a DSW and a RW: (a) Riemann invariants $\lambda_i$; (b) Density $\rho$ distribution; (c) Fluid velocity $v$ distribution.}
\end{center}

Our core objective is to derive an analytical description for the interaction and overtaking process after the DSW catches up with the RW in terms of the initial profile parameters $h_1, h_2$ and $l$.

Adopting the Gurevich-Pitaevskii matching regularization method, scholars have achieved a complete classification for the Riemann problem of Eq.~(\ref{1}) relying on self-similar solutions of Eqs.~(\ref{2.14})~\cite{ck35}, and Eqs.~(\ref{2.3}) and~(\ref{3.3}) subject to $l=0$ belong to one specific case of such classification. Distinct from the dispersive Riemann problem in existing literature, the governing Eqs.~(\ref{2.3}) and~(\ref{3.3}) in this work feature a finite distance $l$ between the discontinuities of $\lambda_{+}$ and $\lambda_{-}$. This feature destroys the self-similar nature of the modulation problem, calling for more general analytical tools.
\vspace{5mm}\\
\textbf{3.2 The mapping to the Euler-Poisson-Darboux equation }
\\

The generalized hodograph method~\cite{ck16,ck37} can be introduced as a method of solving the Whitham equations with number of unknown functions greater than two. Once the collision between the DSW and the RW initiates, two unknown and varying Riemann invariants emerge. Accordingly, we apply the generalized hodograph transformation to reformulate Eqs.~(\ref{2.14}), and the resulting form is presented as follows
\begin{equation}\label{3.5}
x-v_{\mathrm{\it i}}\left(\boldsymbol{\lambda} \right) t=w _{\mathrm{\it i}}\left( \boldsymbol{\lambda} \right) ,\quad i=1,2,3,4,
\end{equation}
where $\boldsymbol{\lambda} = (\lambda_1,\lambda_2,\lambda_3,\lambda_4)$. Differentiation of Eqs.~(\ref{3.5}) with respect to $\lambda_{\mathrm{\it j}}$ gives the relations
\begin{equation}\label{3.6}
-\frac{\partial v_{\mathrm{\it i}}}{\partial \lambda _{\mathrm{\it j}}}t=\frac{\partial w_{\mathrm{\it i}}}{\partial \lambda _{\mathrm{\it j}}},\;\;\;\;i\ne j,
\end{equation}
where the variable $t$ can be given by means of Eqs.~(\ref{3.5})
\begin{equation}\label{3.7}
t=-\frac{w_{\mathrm{\it i}}-w_{\mathrm{\it j}}}{v_{\mathrm{\it i}}-v_{\mathrm{\it j}}},
\end{equation}
so that we arrive at the Tsarev equarions
\begin{equation}\label{3.8}
\frac{1}{w _{\mathrm{\it i}}-w_{\mathrm{\it j}}}\frac{\partial w _{\mathrm{\it i}}}{\partial \lambda _{\mathrm{\it j}}}=\frac{1}{v_{\mathrm{\it i}}-v_{\mathrm{\it j}}}\frac{\partial v_{\mathrm{\it i}}}{\partial \lambda _{\mathrm{\it j}}}.
\end{equation}

It is worth noting that we can provide the Whitham equations with respect to $w_{\mathrm{\it i}}$ that is similar to $v_{\mathrm{\it i}}$
\begin{equation}\label{3.9}
\frac{\partial \lambda _{\mathrm{\it i}}}{\partial \tau}+w_{\mathrm{\it i}}\left( \lambda \right) \frac{\partial \lambda _{\mathrm{\it i}}}{\partial x}=0,\;\;\;\;i=1,2,3,4,
\end{equation}
and the compatibility condition $\partial _{\mathrm{\it t\tau}}\lambda _{\mathrm{\it i}}=\partial _{\mathrm{\it \tau t}}\lambda _{\mathrm{\it i}}.$
Because $v_{\mathrm{\it i}}$ and $w _{\mathrm{\it i}}$ are symmetric, we can seek $w_{\mathrm{\it i}}$ in the form analogous to the Whitham velocities~(\ref{2.15}),
\begin{equation}\label{3.10}
w_{\mathrm{\it i}}=\left( 1-\frac{L}{\partial _{\mathrm{\it i}}L}\partial _{\mathrm{\it i}} \right) W,\quad \partial _i\equiv \partial /\partial \lambda _i.
\end{equation}
Thus, with the help of Eqs.~(\ref{3.10}), Eqs.~(\ref{3.8}) are transformed into the Euler-Poisson-Darboux (EPD) equation for $W(\lambda_i, \lambda_j)$
\begin{equation}\label{3.11}
\frac{\partial ^2W}{\partial \lambda _{\mathrm{\it i}}\partial \lambda _{\mathrm{\it j}}}-\frac{1}{2\left( \lambda _{\mathrm{\it i}}-\lambda _{\mathrm{\it j}} \right)}\left( \frac{\partial W}{\partial \lambda _{\mathrm{\it i}}}-\frac{\partial W}{\partial \lambda _{\mathrm{\it j}}} \right) =0,\;\;\;\;i\ne j.
\end{equation}

The general solution of the EPD equation~(\ref{3.11}) can be represented in the form (see, for instance, \cite{ck40})
\begin{equation}\label{3.12}
W = \int_{a_1}^{\lambda_j}\frac{\phi_1(\lambda)d\lambda}{\sqrt{(\lambda - \lambda_j)(\lambda_i - \lambda)}} + \int_{a_2}^{\lambda_i}\frac{\phi_2(\lambda)d\lambda}{\sqrt{(\lambda - \lambda_j)(\lambda_i - \lambda)}},
\end{equation}
where $\phi_{1,2}(\lambda)$ are arbitrary (generally, complex-valued) functions and $a_{1,2}$ are arbitrary constants (which could be absorbed into $\phi_{1,2}$).

This construction holds for any pair of Riemann invariants, and the EPD equation remains valid even when all four invariants vary. Essentially, the complete integrability of the Hirota-Whitham system guarantees the linearization of the nonlinear modulation equations, which ultimately reduces to the classical linear EPD equation. By converting the free-boundary matching conditions into a Goursat boundary-value problem for Eq.~(\ref{3.11}), the unknown functions $\phi_{1,2}$ can be determined, and the modulation solutions are thereby obtained. It should be noted that the hodograph solution excludes simple wave solutions, which correspond to the degenerate case for the vanishing Jacobian of the hodograph transform $(\lambda_i,\lambda_j)\mapsto(x,t)$\cite{ck1}.
\vspace{5mm}\\
\textbf{3.3 Modulation phase shift }
\\

In the modulated wave, the initial phase shift $Q$ of periodic solutions is no longer an independent constant, but rather a slowly varying function of $x$ and $t$. For this reason, it is more appropriately termed the modulation phase shift. To analyze the modulation phase shift, we reproducing the results reported in Ref.~\cite{ck41}. For the periodic wave~(\ref{2.8}), we present its wavenumber and frequency as follows: 
\begin{equation}\label{3.13}
k = \frac{2\pi}{L} = \frac{\pi\sqrt{(\lambda_1 - \lambda_3)(\lambda_2 - \lambda_4)}}{K(m)}, \quad \omega = kV,
\end{equation}
and the phase shift has the form
\begin{equation}\label{3.14}
\Theta = k(x - Vt - Q).
\end{equation}
After slow modulations of the periodic wave~(\ref{2.8}), the wavenumber and frequency must also satisfy the generalized phase conditions:
\begin{equation}\label{3.15}
k = \frac{\partial \Theta}{\partial x}, \quad \omega = -\frac{\partial \Theta}{\partial t},
\end{equation}
which imply the conservation of waves law
\begin{equation}\label{3.16}
k_t + \omega_x = 0.
\end{equation}

Then, using $\partial_i \omega / \partial_i k = v_i$, we differentiate Eq.~(\ref{3.14}) with respect to $x$ to obtain
\begin{equation}\label{3.17}
\begin{aligned}
\frac{\partial \Theta}{\partial x} &= k + \sum_{i=1}^{4} \frac{\partial \lambda_i}{\partial x} \left( x \frac{\partial k}{\partial \lambda_i} - t \frac{\partial \omega}{\partial \lambda_i} - Q \frac{\partial k}{\partial \lambda_i} - k \frac{\partial Q}{\partial \lambda_i} \right) \\
&= k + \sum_{i=1}^{4} \frac{\partial \lambda_i}{\partial x} \partial_i k \left( x - \frac{\partial_i \omega}{\partial_i k} t - Q - \frac{k}{\partial_i k} \partial_i Q \right) \\
&= k + \sum_{i=1}^{4} \frac{\partial \lambda_i}{\partial x} \partial_i k \left[ x - v_i t - \left( 1 - \frac{L}{\partial_i L} \partial_i \right) Q \right].
\end{aligned}
\end{equation}
Comparing Eq.~(\ref{3.17}) with~(\ref{3.15}), for any pair $i,j$, $i\neq j$, we can conclude that
\begin{equation}\label{3.18}
x - v_n t - \left( 1 - \frac{L}{\partial_n L} \partial_n \right) Q = 0,\quad n=i,j.
\end{equation}
Combining the modulation hodograph solution~(\ref{3.5}) and~(\ref{3.10}) with expression~(\ref{3.18}), we can obtain
\begin{equation}\label{3.19}
\left(1 - \frac{L}{\partial_i L} \partial_i\right) (Q - W) = 0, \quad i=1,2,3,4.
\end{equation}
This means the modulation phase shift $Q(\lambda_1,\lambda_2,\lambda_3,\lambda_4)$ is related to $W(\lambda_1,\lambda_2,\lambda_3,\lambda_4)$, the solution of the boundary value problem for the elliptic partial differential equation~(\ref{3.11}):
\begin{equation}\label{3.20}
Q(\lambda_1,\lambda_2,\lambda_3,\lambda_4) = W(\lambda_1,\lambda_2,\lambda_3,\lambda_4) + C_0 L,
\end{equation}
where $C_0 = \frac{1}{2}$ is a constant (for details, see Ref.~\cite{ck25}). 

For a simple central DSW, only one Riemann invariant varies with $x$ and $t$, while all others remain constant. In this case, the equation reduces to $x - v_m t = 0$. Therefore, we can directly set $Q = L/2$. The trailing dark soliton~(\ref{2.11}) of the DSW is then exactly located at the trailing edge $x^l(t)$ defined by Eqs. ~(\ref{2.17}). Excellent consistency between the numerical and analytical results in the following sections verifies the validity of the modulation phase shift.

\vspace{5mm} \noindent\textbf{4  Modulation structure and spatiotemporal evolution of the co-direction overtaking collision}
\hspace*{\parindent}
\renewcommand{\theequation}{4.\arabic{equation}}\setcounter{equation}{0}

Based on the initial configuration and analytical solution methods for DSW-RW interactions constructed above, we systematically divide the complete dynamic process of DSW catching up with, colliding with and overtaking the RW into three typical chronological stages from the perspective of temporal evolution: pre-collision, during the collision and post-collision, and carries out theoretical analysis and analytical derivation sequentially.
\vspace{5mm}\\
\textbf{4.1 pre-collision ($0<t<t_0$) }
\\

Consistent with the globally stepped initial configuration~(\ref{3.3}) composed of piecewise constant background states, the self-similar modulation solutions characterizing the right-propagating DSW centered at $(0,0)$ generated by the jump of $\lambda_+$ are formulated in terms of Riemann invariants governed by the Whitham modulation equations:
\begin{equation}\label{4.1}
\begin{aligned}
\lambda_1&=0,\quad \lambda_2=1,\quad \lambda_4=h_1,\\
\frac{x}{t}
&=v_3(0,1,\lambda_3,h_1)=V+2\left(\alpha+\beta\left(2\lambda_3+\sum_{i=1}^{4}\lambda_i\right)\right)\frac{(\lambda_2-\lambda_3)(\lambda_3-\lambda_4)}{(\lambda_2-\lambda_3)-(\lambda_2-\lambda_1)\kappa(m)}\\
&=V+2\left(\alpha+\beta(3\lambda_3+1+h_1)\right)\frac{(1-\lambda_3)(\lambda_3-h_1)}{(1-\lambda_3)-(1-h_1)\kappa(m)},
\end{aligned}
\end{equation}
where 
\begin{equation}\label{4.2}
m=\frac{\lambda_3-h_1}{\lambda_3(1-h_1)}.
\end{equation}

From Eqs.~(\ref{4.1}), we can calculate the two boundaries of the DSW. At the leading edge, we set $\lambda_3=1$, i.e., $m=1$, which yields:
\begin{equation}\label{4.3}
x_1^l=\left(-\alpha(2+h_1)-\beta\left(\frac{3}{2}h_1^2+2h_1+4\right)\right)t.
\end{equation}
At the right leading edge $x_1^r$, we set $\lambda_3=h_1$, i.e., $m=0$, which yields:
\begin{equation}\label{4.4}
x_1^r=\left(-\alpha\left(4h_1-\frac{1}{2h_1-1}\right)-\frac{3}{2}\beta\left(8h_1^2-\frac{2}{2h_1-1}-1\right)\right)t.
\end{equation}

We now turn to the right‑propagating RW centered at $x=l$, which can be asymptotically described by similarity solutions of the classical limit equations~(\ref{2.4}) and~(\ref{2.5}):
\begin{equation}\label{4.5}
\lambda_+=1\quad \text{and} \quad
\begin{cases}
\lambda_-=0, & x<x_2^l,\\
\dfrac{x-l}{t}=v_-(\lambda_-,1)=-\alpha(1+3\lambda_-)-\beta\left(\dfrac{3}{2}+3\lambda_-+\dfrac{15}{2}\lambda_-^2\right), & x_2^l\le x\le x_2^r,\\
\lambda_-=h_2, & x>x_2^r.
\end{cases}
\end{equation}

For the boundaries $x_2^{l,r}$, we have
\begin{equation}\label{4.6}
x_2^l=\left(-\alpha-\frac{3}{2}\beta\right)t+l,\quad
x_2^r=\left(-\alpha(1+3h_2)-\beta\left(\frac{3}{2}+3h_2+\frac{15}{2}h_2^2\right)\right)t+l.
\end{equation}

It is worth noting that since the modulation equations reduce to the harmonic wave equations in the harmonic limit, the RW solution~(\ref{4.5}) is also a solution of full modulation system~(\ref{2.14}), namely:
\begin{equation}\label{4.7}
\lambda_3=\lambda_4=h_1,\quad \lambda_2=\lambda_+=1,\quad \lambda_1=\lambda_-(x,t).
\end{equation}
\begin{center}
\includegraphics[scale=0.21]{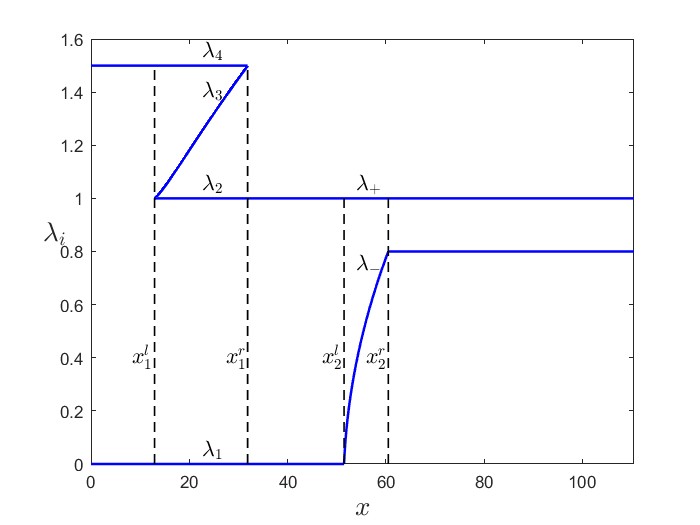}
\includegraphics[scale=0.21]{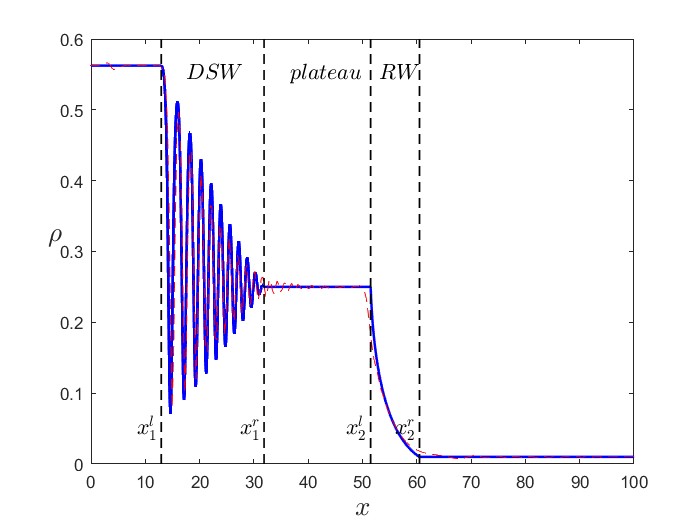}
\includegraphics[scale=0.18]{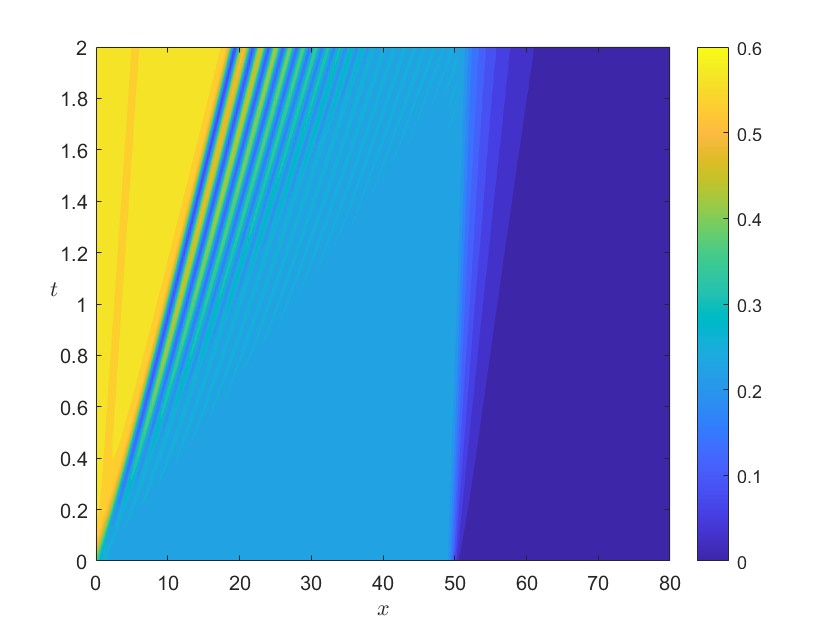}

\vspace{-0.3cm}
{\footnotesize\hspace{-0.2cm}(a)\hspace{4.9cm}(b)\hspace{4.8cm}(c)}

\vspace{-0.2cm}
\flushleft{\footnotesize
\textbf{Fig.~$2$.} (a) Distributions of Riemann invariants, (b) Waveform structure of the density function $\rho$ (red dashed line: numerical simulation; blue solid line: analytical result) and (c) Evolution of the density function $\rho$ before collision. The parameters are taken as $t=1.5$, $h_1 = 1.5$, $h_2 = 0.8$, $l = 50$, $\alpha = 0.5$ and $\beta = -1$.}
\end{center}

Figure~$2$(a) illustrates the schematic evolution of Riemann invariants in the first stage of the co-directional overtaking collision, i.e., the pre-collision. As shown in Fig.~$2$(b), for $0<t<t_0$, the DSW and RW evolve independently without mutual interaction. It can be observed from the evolution in Fig.~$2$(c) that their edges gradually approach each other over time, accompanied by continuous narrowing of the plateau region between the two simple waves. At $t=t_0$, the right edge of the DSW catches up with the left edge of the RW, namely, $x_0=x_1^r(t_0)=x_2^l(t_0)$. Then, by virtue of Eqs.~(\ref{4.4}) and~(\ref{4.6}), we can deduce
\begin{equation}\label{4.8}
t_0=\frac{(1-2h_1)l}{\alpha(8h_1^2-6h_1)+6h_1\beta(4h_1^2-2h_1-1)},\quad
x_0=\frac{2\alpha(8h_1^2-4h_1-1)+3\beta(16h_1^3-8h_1^2-2h_1-1)}{2\alpha(8h_1^2-6h_1)+12h_1\beta(4h_1^2-2h_1-1)}.
\end{equation}
\begin{center}
\includegraphics[scale=0.21]{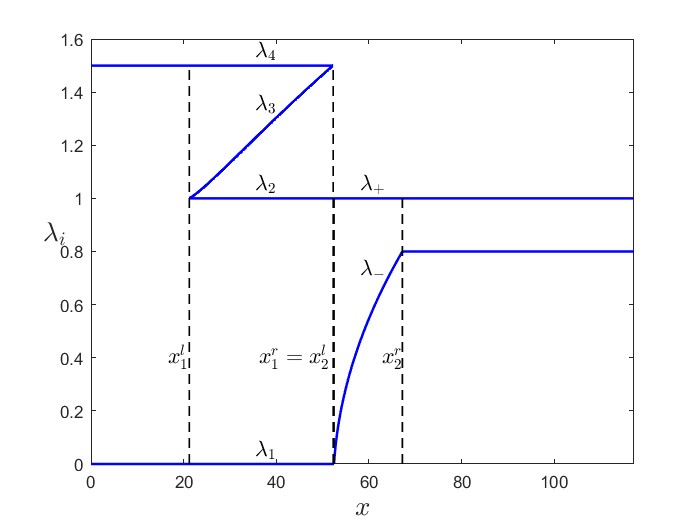}
\includegraphics[scale=0.21]{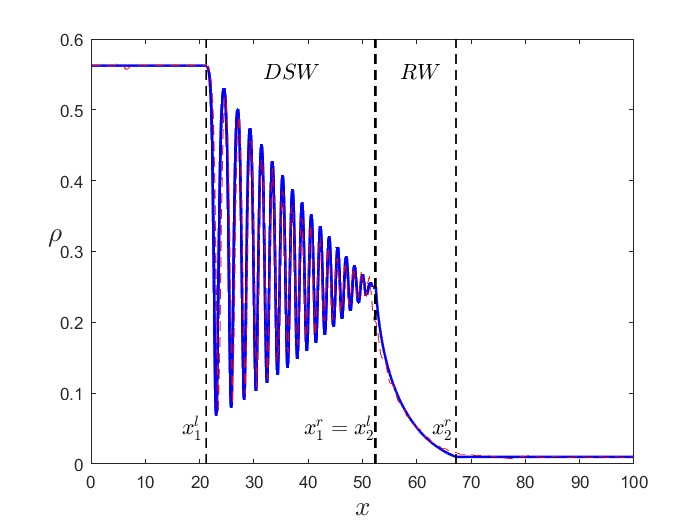}
\includegraphics[scale=0.21]{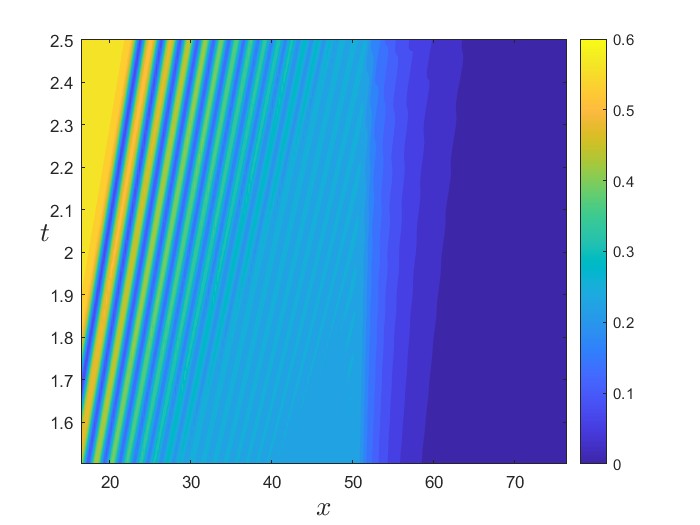}

\vspace{-0.3cm}
{\footnotesize\hspace{-0.2cm}(a)\hspace{5.0cm}(b)\hspace{4.7cm}(c)}

\vspace{-0.2cm}
\flushleft{\footnotesize
\textbf{Fig.~$3$.} (a) Schematic diagram of Riemann invariants, (b) Waveform structure of the density function $\rho$ (red dashed line: numerical simulation; blue solid line: analytical result) and (c) Evolution of the density function $\rho$ at the exact moment of collision. The parameters are taken as $t = t_0=2.4691$, $h_1 = 1.5$, $h_2 = 0.8$, $l = 50$, $\alpha = 0.5$ and $\beta = -1$.}
\end{center}

According to Eqs.~(\ref{4.8}), by choosing appropriate parameters, we can accurately calculate the specific time $t_0$ at which the DSW exactly catches up with the co-propagating RW and begins to interact with it. Fig.~$3$ presents the schematic diagram of the Riemann invariants and the waveform structure of the density $\rho$ at $t_0$, as well as the evolution process of the interaction between the two simple waves near $t_0$.
\vspace{5mm}\\
\textbf{4.2 During the collision ($t_0<t<t_3$)}
\\

At time $t_0$, the domains of the DSW and RW connect precisely, and the two waves collide to form a nonlinear interaction region for $t>t_0$. Using the pre-collision boundary trajectories of the DSW given by Eqs.~(\ref{4.3}) and~(\ref{4.4}) as well as the RW trajectory in Eq.~(\ref{4.6}), we plot the predicted collision paths under the assumption that mutual interaction is absent, as displayed in Fig.~$4$(a). It is worth noting that the overall expansion velocity of the RW is smaller than the propagation speed of the DSW’s right boundary, such that the RW domain is eventually fully engulfed by the DSW domain. As indicated in Fig.~$4$(a), two characteristic times $t_1$ and $t_2$ deserve particular attention: at $t_1$, the right boundary of the DSW catches up with the entire RW domain; at $t_2$, the left boundary of the DSW begins to overtake the RW. For $t_1<t<t_2$, the interaction region coincides with the full RW domain $[x_2^l,x_2^r]$. At $t=t_3$, the DSW overtakes the RW as a whole, and the two wave structures become completely separated eventually.

Throughout the whole interaction process $t_1<t<t_3$, we always have $\lambda_2=1$ and $\lambda_4=h_1$, while the remaining two Riemann invariants $\lambda_1$ and $\lambda_3$ evolve dynamically. Consequently, the modulation solution is no longer self‑similar, and a more general hodograph solution is required. Based on the transformation derived in Section~$3.2$, Eq.~(\ref{3.11}) for $W_{1,3}(\lambda_1,\lambda_3)=W_{1,3}(\lambda_1,1,\lambda_3,h_1)$ can be simplified. Its general solution is parameterized by two arbitrary functions $\boldsymbol{\phi}_{1,2}(\lambda)$, which are determined by the corresponding boundary conditions. These boundary conditions are further derived from the continuous matching conditions of $\lambda_1$ and $\lambda_3$ at the unknown boundaries $x^l(t)$ and $x^r(t)$.
\begin{center}
\includegraphics[scale=0.21]{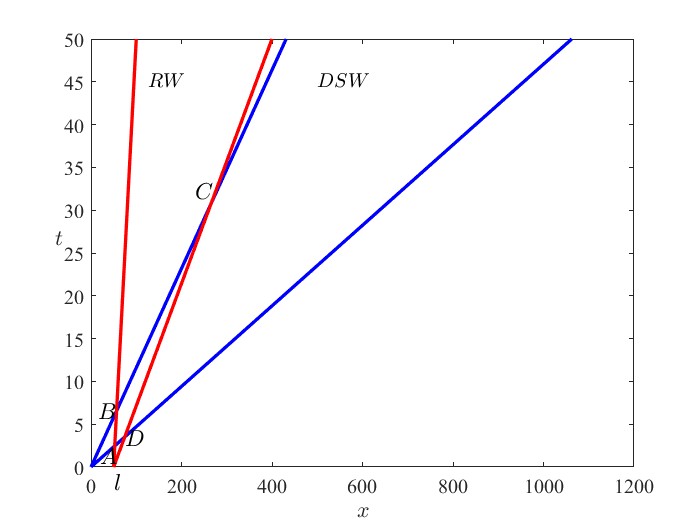}
\includegraphics[scale=0.21]{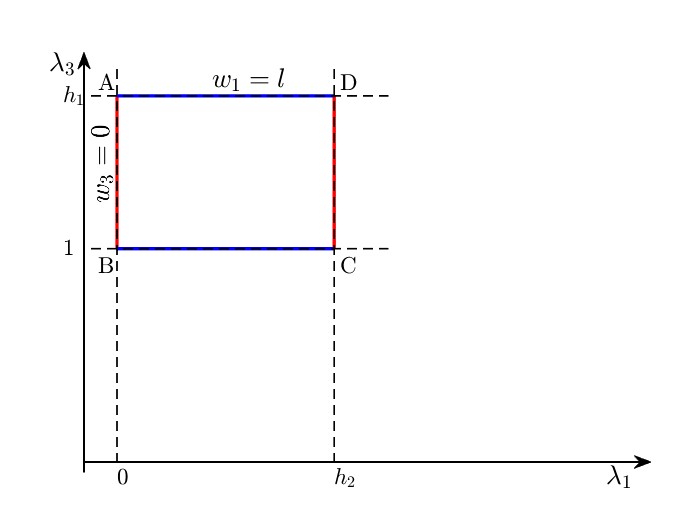}

\vspace{-0.3cm}
{\footnotesize\hspace{-0.2cm}(a)\hspace{4.5cm}(b)}

\vspace{-0.2cm}
\flushleft{\footnotesize
\textbf{Fig.~$4$.} (a) Predicted collision trajectories of DSW-RW, $(x,t)$ plane; (b) Velocity diagram, $(\lambda_1,\lambda_3)$ plane.}
\end{center}

At a segment of the left boundary within the interaction region, namely $x=x_{2}^{l}(t)$, we have
\begin{equation}\label{4.9}
\lambda_1=0,\quad \lambda_2=1,\quad \lambda_3^s(x,t)=\lambda_3\left(\frac{x}{t}\right),\quad\lambda_4=h_1,
\end{equation}
where $\lambda_3$ can be obtained from the self-similar modulation solution~(\ref{4.1}).

At the right boundary $x=x_1^r(t)$ of the interaction region, we have 
\begin{equation}\label{4.10}
\lambda_1=\lambda_1^r\big(x_1^r(t),t\big),\quad \lambda_2=1,\quad \lambda_3=\lambda_4=h_1,
\end{equation}
where $\lambda_1^r\big(x_1^r(t),t\big)= \lambda_1\left(\frac{x-l}{t}\right)$ can be derived from the RW solution given in Eq.~(\ref{4.4}).

Subsequently, we convert the above two nonlinear free-boundary conditions ~(\ref{4.9}) and ~(\ref{4.10}) into boundary conditions for the function $W(\lambda_1,\lambda_3)$ that satisfy the EPD equation
\begin{equation}\label{4.11}
2(\lambda_3-\lambda_1)\partial_{13}^2 W=\partial_3 W-\partial_1 W,\quad \partial _i\equiv \partial /\partial \lambda _i.
\end{equation}
In deriving the boundary conditions for Eq.~(\ref{4.11}), the corresponding Whitham velocities associated with $\lambda_1$ and $\lambda_3$ are given as follows:
\begin{equation}\label{4.12}
\begin{aligned}
&v_1=v_-=\frac{x-l}{t}=-\alpha(3\lambda_1+1)-\frac{3}{2}\beta(5\lambda_1^2+2\lambda_1+1),\\
&v_3=\frac{x}{t}= V+2\left(\alpha+\beta(3\lambda_3+1+h_1)\right)\frac{(1-\lambda_3)(\lambda_3-h_1)}{(1-\lambda_3)-(1-h_1)\kappa(m)},
\end{aligned}
\end{equation}
and the corresponding hodograph solution~(\ref{3.10}) can be written as:
\begin{equation}\label{4.13}
x-v_1 t=w_1,\quad x-v_3 t=w_3.
\end{equation}

Using the boundary condition~(\ref{4.10}) at $x=x_2^l$ and the self-similar form of $v_3$ in expression~(\ref{4.12}), we can derive that
\begin{equation}\label{4.14}
w_3(0,\lambda_3)=0.
\end{equation}
Similarly, according to the boundary condition~(\ref{4.10}) at $x=x_1^r$, we have $\lambda_1=\lambda_-$. Combined with $v_1$ given in expression~(\ref{4.12}), it yields
\begin{equation}\label{4.15}
w_1(\lambda_1,h_1)=l.
\end{equation}

Apparently, as displayed in Fig.~$4$(a), the two simple waves collide and generate the interaction region $ABCD$. This region appears regular in the absence of interaction effects; however, once mutual interaction is taken into account, the quadrilateral $ABCD$ in the $(x,t)$ plane deforms into an irregular curved interaction region with unknown boundary geometry. We therefore map the originally unknown curved interaction region $ABCD$ in the $(x,t)$ plane onto a regular, numerically tractable rectangular domain $ABCD$ defined on the hodograph plane $(\lambda_1,\lambda_3)$ of characteristic velocities. After this transformation, the boundary conditions for the unknown functions $w_{1,3}(\lambda_1,\lambda_3)$ reduce to linear forms on the hodograph plane.

By utilizing the relation between $w_{1,3}(\lambda_1,\lambda_3)$ and $W(\lambda_1,\lambda_3)$, we can deduce the boundary conditions for the function $W(\lambda_1,\lambda_3)$ in Eq.~(\ref{4.11}). From Eq.~(\ref{4.14}), we can further obtain a simple ordinary differential equation:
\begin{equation}\label{4.16}
W(\lambda_1,h_1)-\frac{L(\lambda_1,1,h_1,h_1)}{\partial_1 L(\lambda_1,1,h_1,h_1)}\partial_1 W(\lambda_1,h_1)=l,
\end{equation}
which can be integrated to give the boundary value of the function  $W(\lambda_1,\lambda_3)$ at $\lambda_3=h_1$:
\begin{equation}\label{4.17}
W(\lambda_1,h_1)=C_1 L(\lambda_1,1,h_1,h_1)+l=\frac{C_1}{\sqrt{h_1-\lambda_1}}+l,
\end{equation}
where $C_1$ is an arbitrary integration constant.

Next, from the boundary condition~(\ref{4.15}), we find
\begin{equation}\label{4.18}
W(0,\lambda_3)-\frac{L(0,1,\lambda_3,h_1)}{\partial_3 L(0,1,\lambda_3,h_1)}\partial_3 W(0,\lambda_3)=0,
\end{equation}
whose integration readily yields
\begin{equation}\label{4.19}
W(0,\lambda_3)=C_2 L(0,1,\lambda_3,h_1),
\end{equation}
where $C_2$ is another arbitrary integration constant.

The Eq.~(\ref{4.11}) is a second-order linear hyperbolic partial differential equation. Instead of being well-posed under conventional initial or boundary value conditions, it is solved by means of boundary conditions imposed on two characteristic lines, known as Goursat type characteristic boundary conditions. The corresponding two characteristic lines here are $\lambda_1=0$ and $\lambda_3=h_1$, respectively. Obviously, the two arbitrary functions in the solution expression of the function $W(\lambda_1,\lambda_3)$ should satisfy the boundary conditions~(\ref{4.17}) and~(\ref{4.19}).
Combined with the conclusions in Section~$3.3$, the function $W(0,\lambda_3)$ corresponds to the modulation phase shift of the DSW. Since this DSW is described by the modulation solution of a central simple wave, its modulation phase shift must be zero. Accordingly, setting $C_2=0$, the boundary condition~(\ref{4.19}) reduces to
\begin{equation}\label{4.20}
W(0,\lambda_3)=0,
\end{equation}
which is consistent with the phase-shift constraint condition.

We take $\phi_2(\lambda)\equiv0,\ a=0$, then
\begin{equation}\label{4.21}
W=\int_{0}^{\lambda_1}\frac{\phi_1(\lambda)d\lambda}{\sqrt{(\lambda_3-\lambda)(\lambda_1-\lambda)}}.
\end{equation}
We therefore only need to determine $\phi_1$ and $C_1$ that satisfy the two boundary conditions~(\ref{4.17}) and~(\ref{4.19}). Substituting Eq.~(\ref{4.20}) into the boundary condition~(\ref{4.17}) yields:
\begin{equation}\label{4.22}
\int_{0}^{\lambda_1}\frac{\phi_1(\lambda)d\lambda}{\sqrt{(h_1-\lambda)(\lambda_1-\lambda)}}=\frac{C_1}{\sqrt{h_1-\lambda_1}}+l.
\end{equation}
It takes the form of an Abel integral equation~\cite{ck42}, from which we can obtain its solution as follows:

\begin{equation}\label{4.23}
\phi_1(\lambda)=\frac{1}{\pi\sqrt{\lambda}}\left(C_1\sqrt{\frac{h_1}{h_1-\lambda}}+l\sqrt{h_1-\lambda}\right).
\end{equation}
It is readily verified that Eqs.~(\ref{4.21}) and~(\ref{4.23}) satisfy condition~(\ref{4.20}) solely under the constraint $\phi_1(0)=0$, which yields $C_1=-l\sqrt{h_1}$ and consequently
\begin{equation}\label{4.24}
W(\lambda_1,\lambda_3)=-\frac{l}{\pi}\int_{0}^{\lambda_1}\frac{\sqrt{\lambda}\,d\lambda}{\sqrt{(\lambda_3-\lambda)(\lambda_1-\lambda)(h_1-\lambda)}}
=\frac{2lh_1}{\pi\sqrt{(h_1-\lambda_1)\lambda_3}}\big(\Pi_1(n,s)-K(s)\big),
\end{equation}
where $\Pi_1(n,s)$ is the complete elliptic integral of the third kind and
\begin{equation}\label{4.25}
s=\frac{\lambda_1(h_1-\lambda_3)}{\lambda_3(h_1-\lambda_1)},\quad n=\frac{\lambda_1}{\lambda_1-h_1}.
\end{equation}

Therefore, the modulation solution characterizing the interaction region of co-propagating DSW and RW is given by the following expressions
\begin{equation}\label{4.26}
\lambda_2=1,\quad \lambda_4=h_1,\quad
x-v_{1,3}(\lambda_1,1,\lambda_3,h_1)t=\left(1-\frac{L}{\partial_{1,3}L}\partial_{1,3}\right)W(\lambda_1,\lambda_3).
\end{equation}

Three critical moments divide the whole interaction process of co-propagating DSW and RW into three stages:

For $t_0<t<t_1$, the interaction region is $[x_2^l,x_1^r]$ (see Fig.~$5$). In this stage, the harmonic edge of the DSW catches up with the left edge of the RW and continuously pursues its other edge. At $t=t_1$, the harmonic edge of the DSW exactly reaches the right edge of the RW with $x_1^r(t_1)=x_2^r(t_1)$, meaning the entire RW region is fully overtaken. From the sketch of Riemann invariants in Fig.~$6$(a), it can be seen that the Riemann invariants satisfy $\lambda_2=1$ and $\lambda_4=h_1$ constantly, with $\lambda_3=h_1$ and $\lambda_1=h_2$ holding simultaneously. Substituting $\lambda_3=h_1$ and $\lambda_1=h_2$ into the hodograph solution~(\ref{4.26}), we obtain after algebraic manipulation that:
\begin{equation}\label{4.27}
t_1 = \frac{(2h_1-1)l}{2\sqrt{h_1(h_1-h_2)}\left[\left(-3+4h_1-h_2\right)\alpha+
3\beta\left(4h_1^2+2h_1(-1+h_2)-(1+h_2)^2\right)\right]}.
\end{equation}
\begin{center}
\includegraphics[scale=0.21]{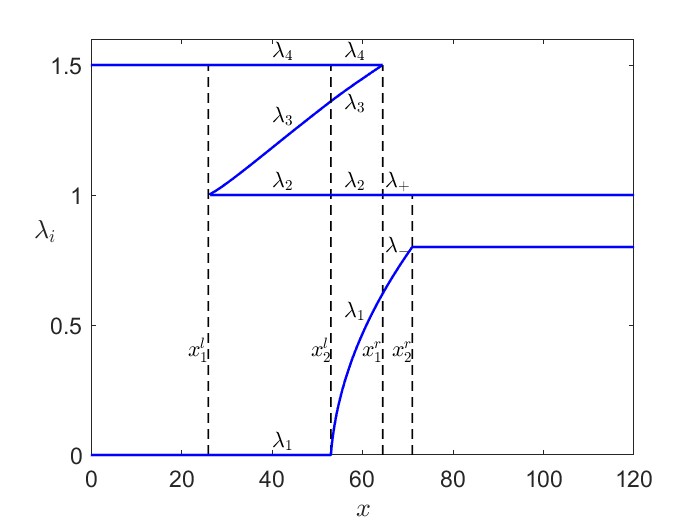}
\includegraphics[scale=0.21]{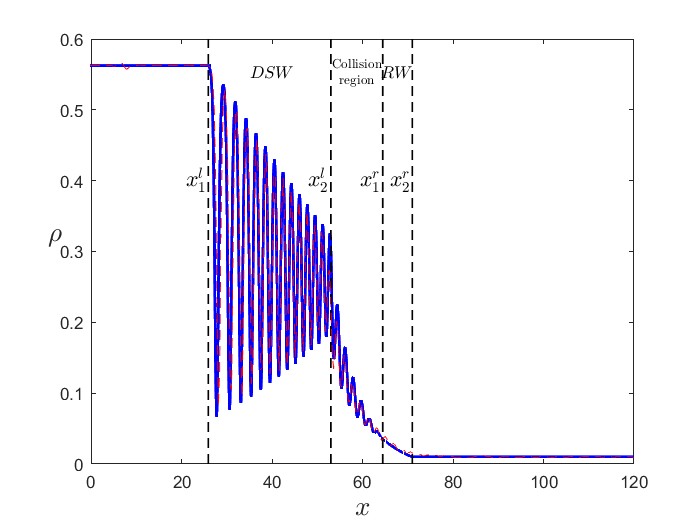}
\includegraphics[scale=0.18]{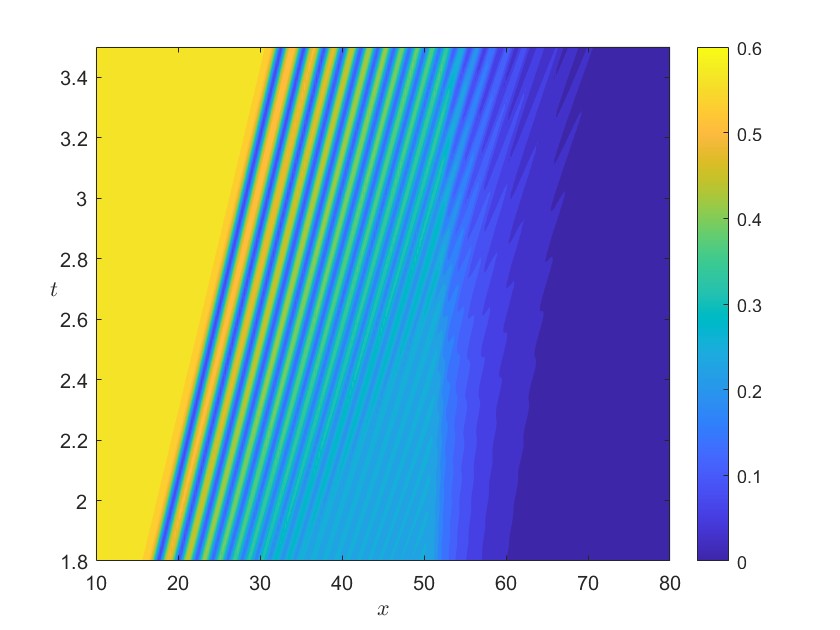}

\vspace{-0.3cm}
{\footnotesize\hspace{-0.2cm}(a)\hspace{4.9cm}(b)\hspace{4.8cm}(c)}

\vspace{-0.2cm}
\flushleft{\footnotesize
\textbf{Fig.~$5$.} First stage of DSW-RW interaction for $t_0<t<t_1$:
(a) Schematic distribution of Riemann invariants;
(b) Waveform structure of the density function $\rho$ (red dashed line: numerical simulation; blue solid line: analytical solution);
(c) Evolution of the density function $\rho$ around time $t$.
Parameters: $t = 3$, $h_1 = 1.5$, $h_2 = 0.8$, $l = 50$, $\alpha = 0.5$, $\beta = -1$.}
\end{center}

For $t_1<t<t_2$, the two waves overlap maximally, and the interaction region covers the whole RW region $[x_2^l,x_2^r]$ (see Fig.~$6$). Although the harmonic edge of the DSW has overtaken the entire RW region, its soliton edge has not yet reached the left edge of the RW. At $t=t_2$, the soliton edge of the DSW precisely catches up with the left edge of the RW satisfying $x_1^l(t_2)=x_2^l(t_2)$, after which the spatial overlap of the two waves gradually decreases. Correspondingly, as seen from the Riemann invariants in Fig.~$6$ (c), the matching conditions reduce to $\lambda_1=0$ and $\lambda_3=1$, and the expression for $t_2$ is given as follows:
\begin{equation}\label{4.28}
t_2 = -\frac{l}{2\sqrt{h_1}\left(\alpha+\beta(2+h_1)\right)}.
\end{equation}

\begin{center}
\includegraphics[scale=0.21]{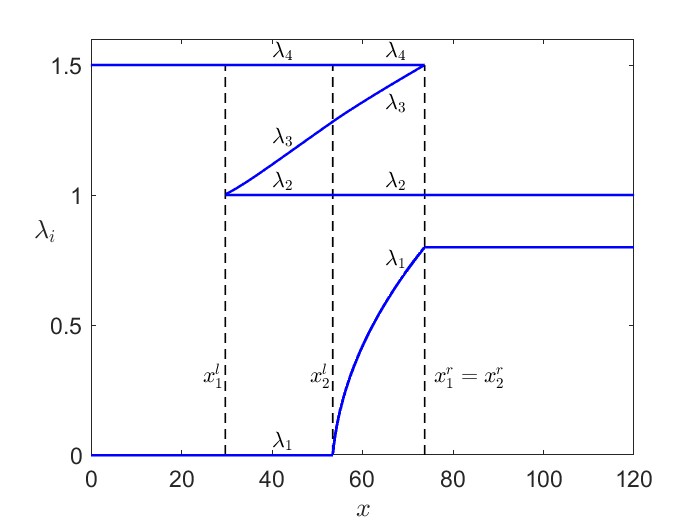}
\includegraphics[scale=0.21]{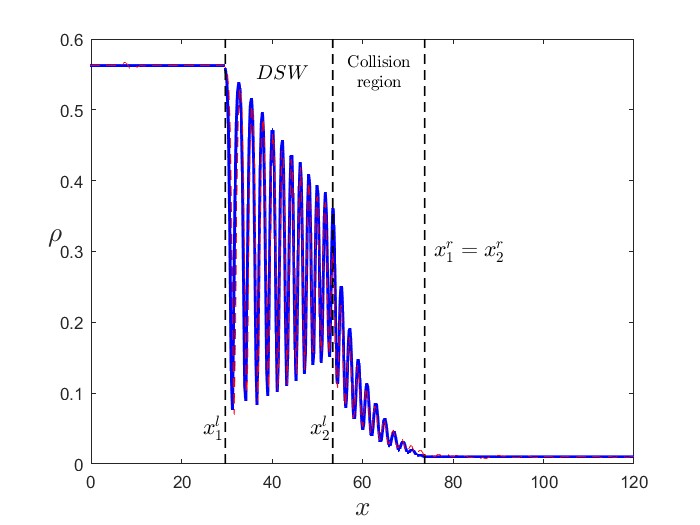}
\includegraphics[scale=0.18]{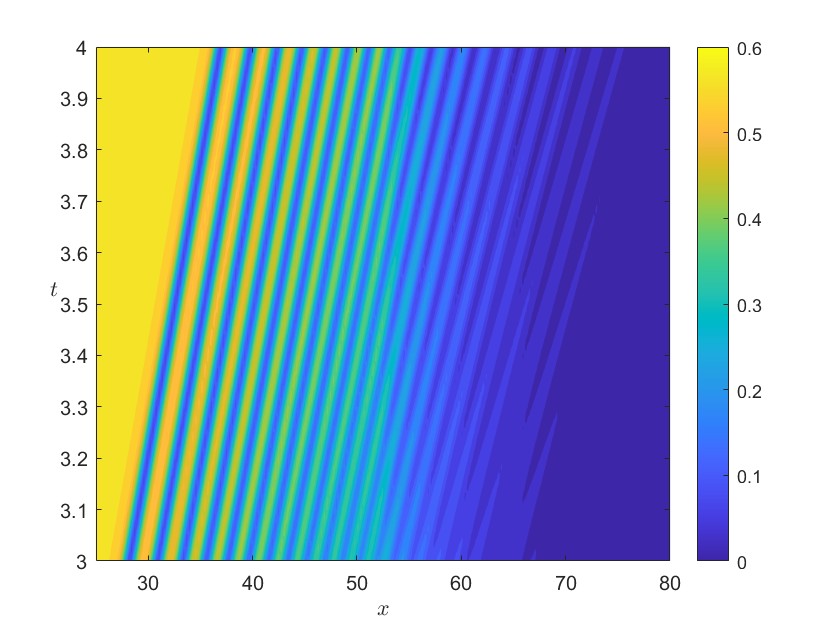}
\vspace{0.1cm}{\footnotesize\hspace{0.1cm}(a)\hspace{4.8cm}(b)\hspace{4.6cm}(c)}

\includegraphics[scale=0.21]{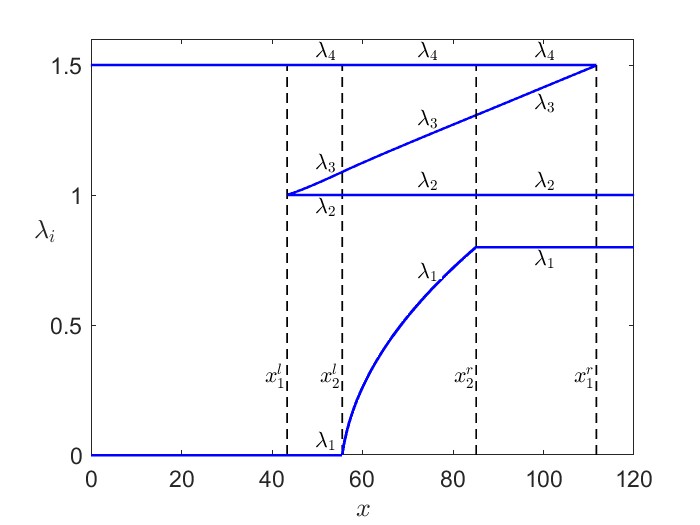}
\includegraphics[scale=0.21]{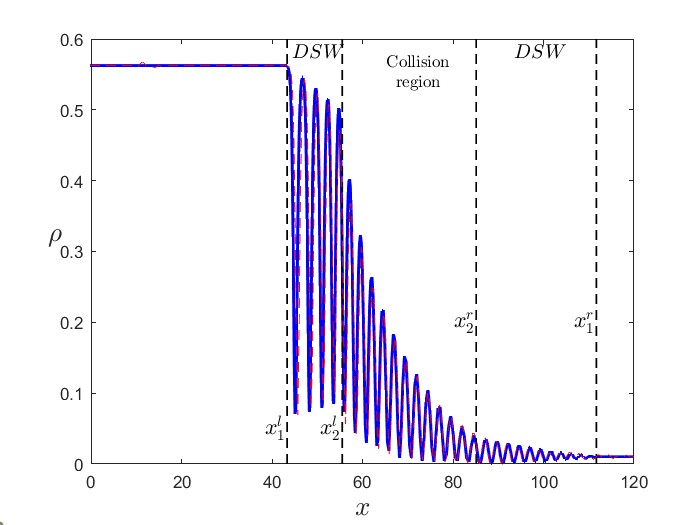}
\includegraphics[scale=0.18]{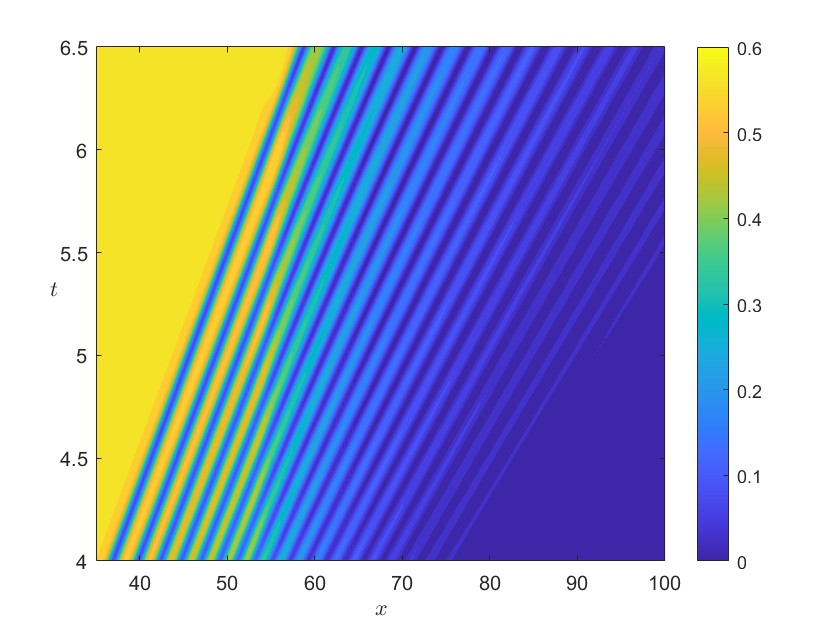}
\vspace{0.1cm}{\footnotesize\hspace{0.15cm}(d)\hspace{4.8cm}(e)\hspace{4.6cm}(f)}

\includegraphics[scale=0.21]{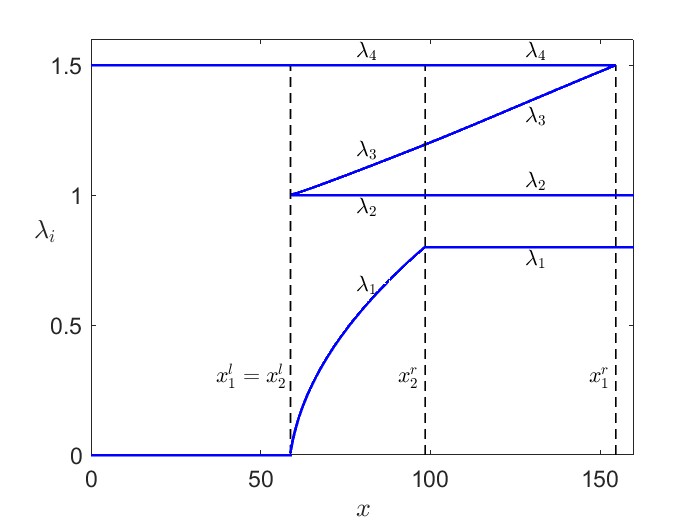}
\includegraphics[scale=0.21]{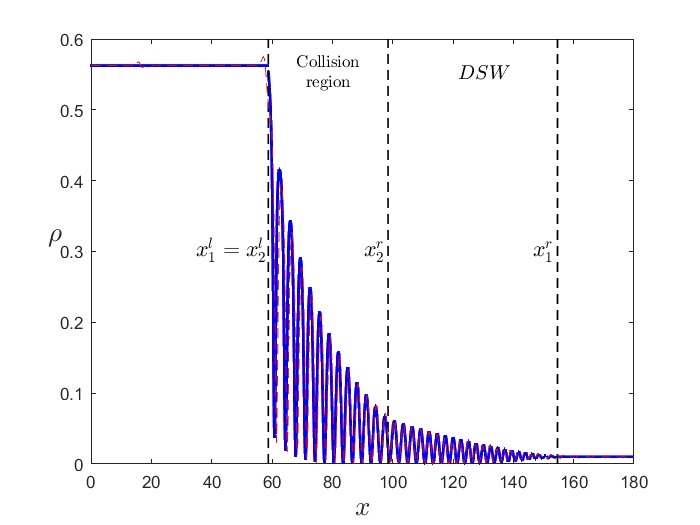}
\includegraphics[scale=0.18]{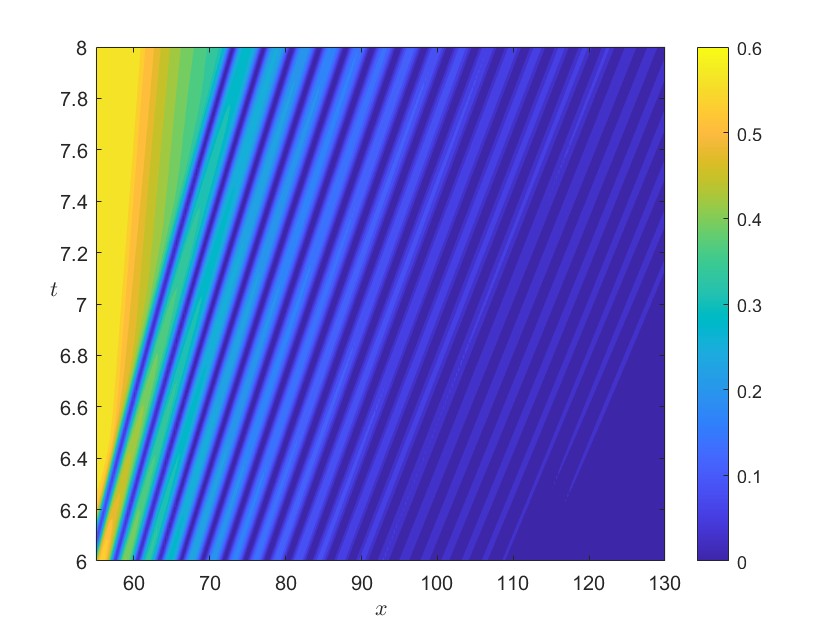}
\vspace{0.1cm}{\footnotesize\hspace{0.1cm}(g)\hspace{4.8cm}(h)\hspace{4.6cm}(i)}
\vspace{-0.2cm}
\flushleft{\footnotesize
\textbf{Fig.~$6$.} Second stage of DSW-RW interaction for $t_1\le t\le t_2$:
the first column shows schematic distributions of Riemann invariants,
the second column presents waveform structures of the density function $\rho$ (red dashed line: numerical simulation; blue solid line: analytical solution),
and the third column displays evolutions of the density function $\rho$ near $t_1$, $t$ and $t_2$, respectively.
Parameters: $t_1=3.393$, $t=5$, $t_2=6.80$, $h_1 = 1.5$, $h_2 = 0.8$, $l = 50$, $\alpha = 0.5$, $\beta = -1$.}
\end{center}

For $t_2<t<t_3$, the soliton edge of the DSW has overtaken the left edge of the RW, and the spatial overlap of the two waves continues to decline, with the interaction region changing to $[x_1^l,x_2^r]$ (see Fig.~$7$). At $t=t_3$, the soliton edge of the DSW exactly reaches the right edge of the RW, i.e., $x_1^l(t_3)=x_2^r(t_3)$, marking the completion of the whole interaction. 
\begin{center}
\includegraphics[scale=0.21]{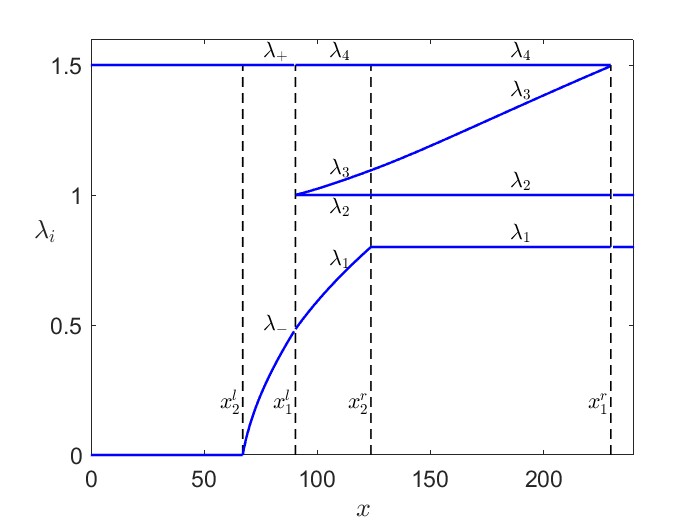}
\includegraphics[scale=0.21]{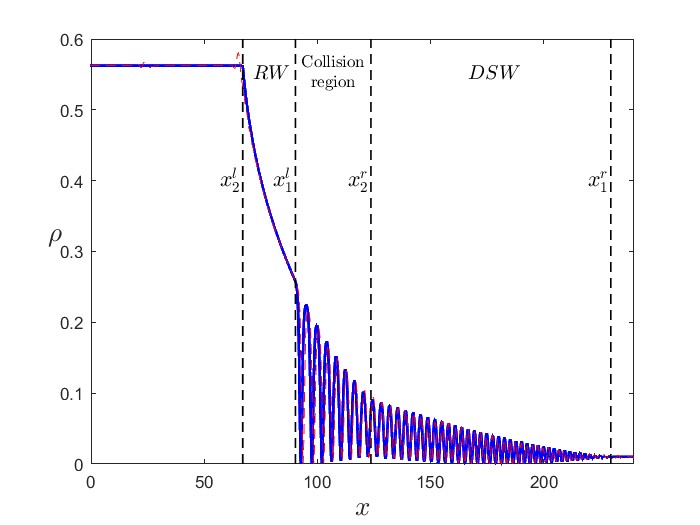}
\includegraphics[scale=0.18]{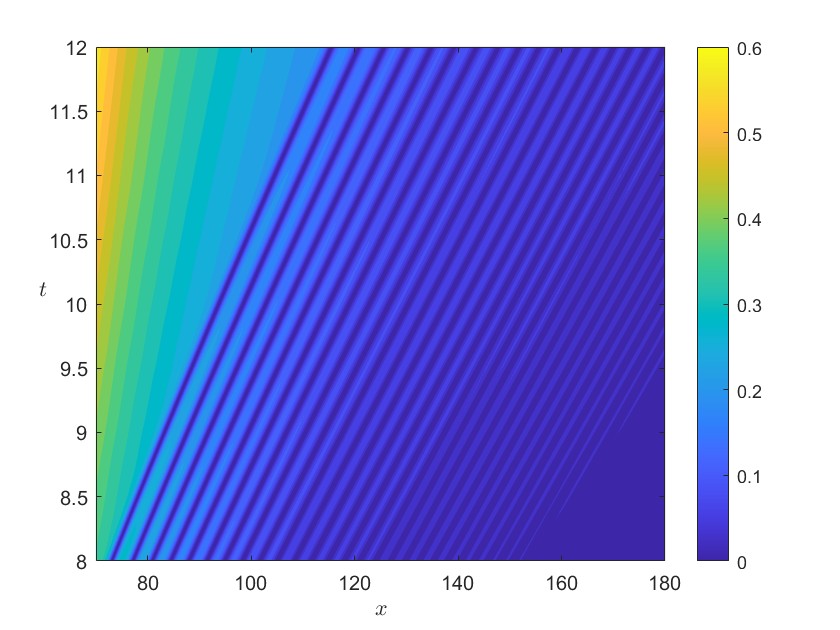}

\vspace{-0.3cm}
{\footnotesize\hspace{-0.2cm}(a)\hspace{4.9cm}(b)\hspace{4.8cm}(c)}

\vspace{-0.2cm}
\flushleft{\footnotesize
\textbf{Fig.~$7$.} Third stage of DSW-RW interaction for $t_2<t<t_3$:
(a) Schematic distribution of Riemann invariants;
(b) Waveform structure of the density function $\rho$ (red dashed line: numerical simulation; blue solid line: analytical solution);
(c) Evolution of the density function $\rho$ around time t .
Parameters: $t = 10$, $h_1 = 1.5$, $h_2 = 0.8$, $l = 50$, $\alpha = 0.5$, $\beta = -1$.}
\end{center}

Clearly, $\lambda_3=1$ remains constant along the left edge of the interaction region, while $\lambda_1$ increases gradually until the interaction region vanishes, where $\lambda_1=h_2$ holds constantly. By inserting $\lambda_3=1$ and $\lambda_1=h_2$ into Eqs.~(\ref{4.26}), we derive the expression for the collision-ending time $t_3$ as follows:
\begin{equation}\label{4.29}
t_3=\frac{lE(r)}{\pi(h_2-1)\sqrt{h_1-h_2}\big(\alpha+\beta(2+3h_2+h_1)\big)},
\end{equation}
where 
$$r=-\frac{h_2(-1+h_1)}{h_2-h_1}.$$
The corresponding coordinate $x_1^l(t_3)$ is given by
\begin{equation}\label{4.30}
x_1^l(t_3)=\left(-\alpha(3h_2+h_1)-\frac{3}{2}\beta(5h_2^2+2h_2+h_1^2)\right)t_3+G(1),
\end{equation}
where $G(1) = w(h_2,1).$
\\
\vspace{5mm}\\
\textbf{4.3 Post-collision ($t>t_3$) }
\\

At $t=t_3$, the DSW and RW are completely separated and resume propagation as two simple waves. After separation, the DSW and RW continue to propagate in the same direction, with the overall propagation speed of the DSW larger than that of the RW.
\vspace{3mm}\\
\textbf{\uppercase\expandafter{\romannumeral1}  Separated DSW}
\vspace{3mm}

After separation, the modulated solution describing the separated DSW is given by three constant Riemann invariants: 
\begin{equation}\label{4.31}
\lambda_1=h_2,\quad \lambda_2=1,\quad \lambda_4=h_1.
\end{equation}
Meanwhile, for the remaining Riemann invariant $\lambda_3$, from the hodograph solution~(\ref{4.26}) we can obtain a simple wave modulation solution 
\begin{equation}\label{4.32}
\begin{aligned}
x&=v_3(h_2,1,\lambda_3,h_1)t+G(\lambda_3)\\
&=-\alpha(h_1+1+h_2+\lambda_3)-\beta \left(\frac{3}{2} (h_1+1+h_2+\lambda_3)^2 - 2 (h_2+ h_1\lambda_3 + h_2 h_1+  \lambda_3 + h_2 \lambda_3 +h_1)\right)+\\
&\quad2\left(\alpha+\beta(3\lambda_3+1+h_1+h_2)\right)\frac{(1-\lambda_3)(\lambda_3-h_1)}{(1-\lambda_3)-(1-h_1)\kappa(m)}+G(\lambda_3),
\end{aligned}
\end{equation}
where 
$$m=\frac{(h_2-1)(\lambda_3-h_1)}{(h_2-\lambda_3)(1-h_1)}$$
and the function $G(\xi)$ is found as
\begin{equation}\label{4.33}
\begin{aligned}
G(\xi)&=w_3(h_2,\xi)=(1-\frac{L(h_2,1,\xi,h_1)}{\partial_3 L(h_2,1,\xi,h_1)}\partial_3) W(h_2,\xi)\\
&=\frac{2l}{\pi\sqrt{\xi(h_2-h_1)}}\left(h_1\Pi_1(p,s)+\frac{h_1(h_1-1)(\xi-h_2)K(s)\kappa(y)+\xi(\xi-1)(h_2-h_1)E(s)}{(\xi-h_2)(1-h_1)\kappa(y)+(\xi-1)}\right),
\end{aligned}
\end{equation}
where
\begin{equation}\label{4.34}
p=\frac{h_2}{h_2-h_1},\quad s=\frac{h_2(\xi-h_1)}{\xi(h_2-h_1)},\quad y=\frac{(h_2-1)(\xi-h_1)}{(\xi-h_2)(h_1-1)}.
\end{equation}
The function $G(\xi)$~(\ref{4.33}) and expression~(\ref{4.34}) are derived from Eqs.~(\ref{4.24}) and~(\ref{4.25}) by setting $\lambda_1=h_2$ and $\lambda_3=\xi$. Consequently, as a result of the interaction, the DSW is no longer described by a self-similar modulation solution in the form of an expanding central fan, but evolves into a more general simple wave solution corresponding to the initial value problem of Eq.~(\ref{1}):
\begin{equation}\label{4.35}
\lambda_-(x,0)=h_2,\quad \lambda_+(x,0)=G^{-1}(x),
\end{equation}
being noted that $G^{-1}(x)$ is the inverse function of $x=G(\lambda_+)$.

The boundaries $x_1^l$ and $x_1^r$ of the separated DSW are found in the modulation solution~(\ref{4.32}) by setting $\lambda_3=1$, (i.e., $m=1$) and $\lambda_3=h_1$, (i.e., $m=0$), respectively:
\begin{equation}\label{4.36}
\begin{aligned}
&x_1^l=\left(-\alpha(h_1+h_2+2)-\beta\left(\frac{3}{2}(h_1^2+h_2^2)+(h_1+2)(h_2+2)\right)\right)t+G(1),\\
&x_1^r=\left(-\alpha\left(4h_1-\frac{(h_2-1)^2}{2h_1-h_2-1}\right)-\beta\left(-12h_1^2+\frac{3}{2}(h_2-1)^2\frac{2h_1+h_2+1}{2h_1-h_2-1}\right)\right)t+G(h_1).
\end{aligned}
\end{equation}
\\
\vspace{3mm}\\
\textbf{\uppercase\expandafter{\romannumeral2} Separated RW}
\vspace{3mm}

The solution for the separated RW is derived from the hodograph modulation solution by setting $\lambda_4=\lambda_+=h_1$, $\lambda_3=\lambda_2=1$, $\lambda_1=\lambda_-$, together with $v_1(\lambda_1,\lambda_2,\lambda_3,\lambda_4)=v_-(\lambda_1,\lambda_4)$. This yields
\begin{equation}\label{4.37}
\begin{aligned}
\lambda_+=h_1,\quad 
x&=v_-(\lambda_-,h_1)t+H(\lambda_-)\\
&=\left(-\alpha(h_1+3h_2)-\beta\left(\frac{3}{2}h_1^2+3h_1\lambda_-+\frac{15}{2}\lambda_-^2\right)\right)t+H(\lambda_-),
\end{aligned}
\end{equation}
where the function $H(\xi)$ has the form
\begin{equation}\label{4.38}
H(\xi)=W_1(\xi,h_1)=\frac{2l}{\pi\sqrt{h_1-\xi}}\big(h_1\Pi_1(z,r)-h_1K(r)+E(r)\big),
\end{equation}
where
\begin{equation}\label{4.39}
z=\frac{\xi}{\xi-h_1},\quad r=\frac{\xi(h_1-1)}{h_1-\xi}.
\end{equation}

Similar to the separated DSW, the separated RW is no longer described by a central fan solution, but by a more general simple-wave solution of the zero-phase solution system with initial conditions $\lambda_+=h_1$ and $\lambda_-(x,0)$ given by the inverse function of $H(\lambda_-)$.

The boundaries of the separated RW are given by the formulaes
\begin{equation}\label{4.40}
\begin{aligned}
&x_2^l=\left(-\alpha h_1-\frac{3}{2}\beta h_1^2\right)t+H(0),\\
&x_2^r=\left(-\alpha(h_1+3h_2)-\beta\left(\frac{3}{2}h_1^2+3h_1h_2+\frac{15}{2}h_2^2\right)\right)t+H(h_2).
\end{aligned}
\end{equation}
\begin{center}
\includegraphics[scale=0.21]{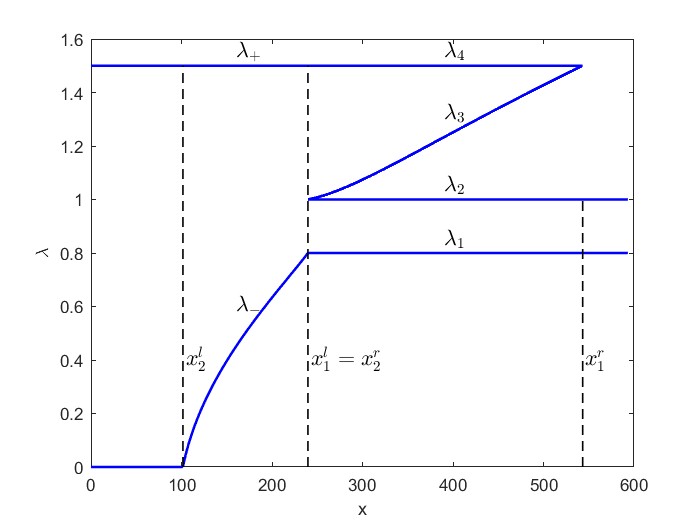}
\includegraphics[scale=0.17]{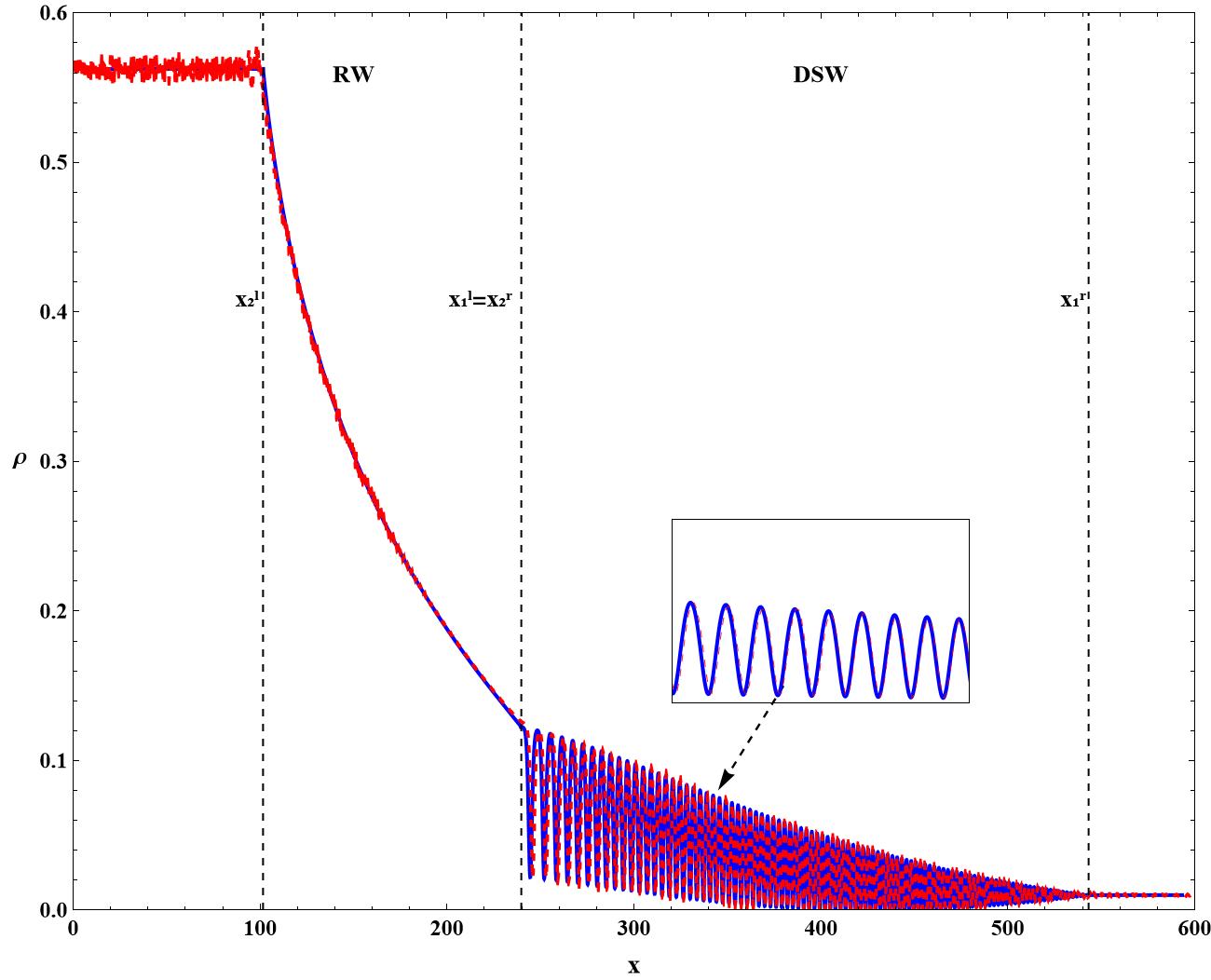}
\includegraphics[scale=0.18]{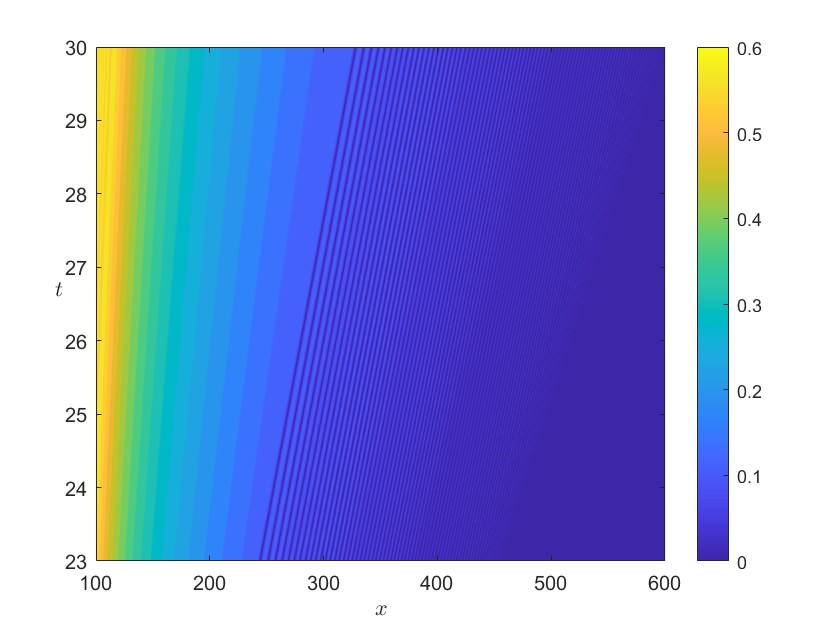}

\vspace{-0.2cm}
{\footnotesize\hspace{0cm}(a)\hspace{4.9cm}(b)\hspace{4.55cm}(c)}

\vspace{-0.2cm}
\flushleft{\footnotesize
\textbf{Fig.~$8$.} After the termination of DSW-RW interaction:
(a) Schematic distribution of Riemann invariants;
(b) Waveform structure of the density function $\rho$ (red dashed line: numerical simulation; blue solid line: analytical solution);
(c) Evolution of the density function $\rho$ near $t$.
Parameters: $t = t_3=23.14$, $h_1 = 1.5$, $h_2 = 0.8$, $l = 50$, $\alpha = 0.5$, $\beta = -1$.}
\end{center}

It can be clearly seen in Fig.~$8$(c) that two separated simple waves evolve after $t_3$, with the RW on the left and the DSW on the right, both propagating rightward synchronously. As indicated in Fig.~$8$(b), owing to the mutual interaction between the DSW and RW, the trailing dark soliton of the DSW suffers evident amplitude attenuation after collision; meanwhile, the overall profile of the RW rises accompanied by an increase in its peak amplitude.
\begin{center}
\includegraphics[scale=0.26]{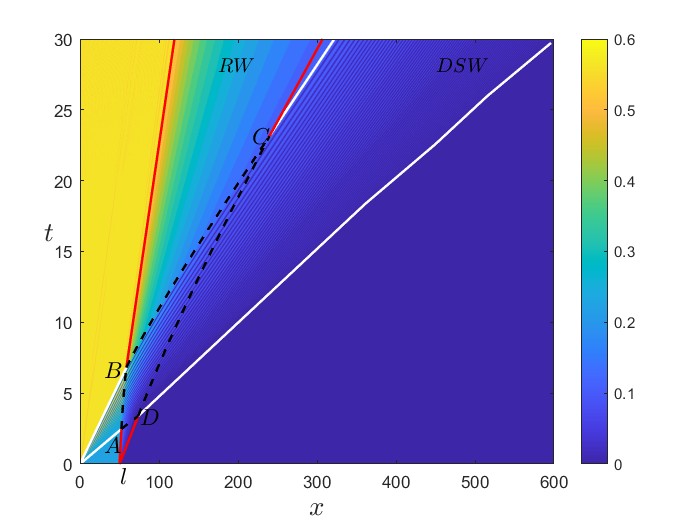}

\vspace{-0.2cm}
\flushleft{\footnotesize
\textbf{Fig.~$9$.} Spatiotemporal evolution diagram and edge trajectory diagram for the entire overtaking collision between DSW and RW.}
\end{center}

Finally, the edge motion laws at each stage of DSW-RW overtaking collision are derived from Eqs.~(\ref{4.26}). Using these laws, we revise the theoretically predicted edge trajectories on the $xt$-plane in Fig.~$4$(a), and the corrected results are presented in Fig.~$9$, which are in good agreement with the full evolution from numerical simulations. Due to the DSW-RW interaction, the originally predicted edge trajectories deviate markedly from the actual evolution, and the collision terminates earlier than the theoretical forecast.

\vspace{5mm}\noindent\textbf{5  Conclusions}\\
\hspace*{\parindent}

Within the framework of Eq.~(\ref{1}), we investigated the complete dynamical characteristics of overtaking collision between a DSW and a co-propagating RW. Firstly, individual RW and DSW were characterized via zero-phase and single-phase modulation systems and periodic solutions. Global initial configurations of Riemann invariants $\lambda_{\pm}$ were established subject to respective transition conditions of the two waves. Secondly, the generalized hodograph transformation was applied to derive the EPD equation governing $W(\lambda_i,\lambda_j)$, and its general solution was formulated to facilitate analytical analysis of nonlinear interaction zones. Finally, the whole overtaking process was divided into three stages: pre-collision, during the collision and post-collision.

In the pre-collision stage, DSW and RW were spatially separated and propagated steadily rightward. Constant Riemann invariants dominated uniform zero-phase modulation on flat regions. The two simple waves gradually approached each other due to velocity discrepancy with independent uncoupled modulation fields.

During collision, mutual interaction emerged. Boundary conversions from $\lambda_i$ to $w(\lambda_1,\lambda_3)$ and further to $W(\lambda_1,\lambda_3)$ were performed to solve the EPD equation and obtain modulated solutions for the interaction region. Intense field coupling and spatial superposition occurred within the time interval $t_1<t<t_2$ driven by edge velocity differences.

In the post-collision stage, the waves separated completely and coupling effects vanished. DSW and RW resumed independent rightward propagation, marking the completion of the overtaking collision dynamics.

In fact, the research framework concerning the nonlinear interaction between DSWs and RWs in the defocusing Hirota equation still possesses considerable expandable potential, and the coupled collision behaviors of two nonlinear waves remain to be further explored with abundant innovative contents. This work only investigates the overtaking collision mechanism between a DSW and a RW within the Whitham modulation framework, while the relevant theoretical analyses can be further extended to more typical nonlinear wave collision scenarios. On this basis, researchers can systematically study the dynamics of head-on and overtaking collisions between DSWs in the defocusing Hirota equation. Such complex collision processes of double DSWs generally involve the multiphase Whitham theoretical framework, which further enriches the dynamic mechanism of nonlinear wave collisions. The above extended investigations effectively improve the nonlinear wave theoretical system of the defocusing Hirota equation and provide more comprehensive and solid theoretical support for the nonlinear wave dynamics in photon hydrodynamics.

\vspace{5mm}\noindent\textbf{Acknowledgments}\\
\hspace*{\parindent}

We express our sincere thanks to each member of our discussion group for their suggestions. This work has been supported by the National Natural Science Foundation of China under Grant No. 12575005, the Shanxi Province Science Foundation under Grant No. 202303021221031, and the Research Project Supported by Shanxi Scholarship Council of China under Grant No. 2024-033.

\end{document}